\documentstyle[12pt]{article}

\def\beq{\begin{equation}}
\def\eeq{\end{equation}}

\newcommand{\sect}[1]{\setcounter{equation}{0}\section{#1}}

\newcommand{\EQ}{\begin{equation}}
\newcommand{\EN}{\end{equation}}
\newcommand{\bea}{\begin{eqnarray}}
\newcommand{\ena}{\end{eqnarray}}

\newcommand{\e}{\epsilon}

\newcommand{\lbl}{\label}
\newcommand{\fslu}{\not\mbox{\hspace{-1.5mm}}}

\def\thinspace{\kern .16667em}

\def\Dir{\nabla\kern-2ex\Big{/}}

\def\reali{{\hbox{\s@ l\kern-.5ex R}}}
\def\naturali{{\hbox{\s@ l\kern-.5ex N}}}
\def\uno{{\hbox{ 1\kern-.8mm l}}}
 
\def\part{\partial}

\def\aa{\alpha}
\def\bb{\beta}
\def\cc{\chi}
\def\dd{\delta}
\def\ee{\epsilon}
\def\vf{\varphi}
\def\GG{\Gamma}
\def\pp{\psi}
\def\PP{\Psi}
\def\pb{\bar\psi}
\def\PB{\bar\Psi}
\def\rr{\rho}

\newcommand{\mat}[4]{\left( 
                     \begin{array}{cc}
                     {#1} & {#2} \\
                     {#3} & {#4} 
                     \end{array}
                     \right)
                    }
\newcommand{\vett}[2]{\left(
                      \begin{array}{c}
                     {#1} \\
                     {#2} 
                     \end{array}
                     \right)
                    }
\newcommand{\vet}[2]{\left(
                     \begin{array}{cc}
                     {#1} &  {#2} 
                     \end{array}
                     \right)
                    }
\newcommand{\ft}[3]{\int {d^{#1}{#2}\over (2\pi)^{#1}} ~ e^{i {#2}\cdot{#3}} }

\def\rhm{\rho_-}
\def\rhp{\rho_+}
\def\sgm{\sigma_-}
\def\sgp{\sigma_+}

\def\rd{\sqrt{2}}
\def\usrd{{1\over\sqrt{2}}}
\def\dxy{\delta^2(x-y)}
\def\dij{\delta^{ij}}
\def\dsi{\partial_{x^-}}
\def\dta{\partial_{x^+}}

\newcommand\modu[1]{|{#1}|}

\newcommand\psiind[3]{\psi^{#1 #2}_{#3}}
\newcommand\psibind[3]{{\bar\psi}^{#1 #2}_{#3}}
\newcommand\psiindd[2]{\psi^{#1 #2}}
\newcommand\psibindd[2]{{\bar\psi}^{#1 #2}}

\def\pbaj{\psibindd{A}{j}}
\def\pai{\psiindd{A}{i}}

\def\paip{\psiind{A}{i}{+}}
\def\pajp{\psiind{A}{j}{+}}
\def\pbip{\psiind{B}{i}{+}}
\def\pbjp{\psiind{B}{j}{+}}
\def\paim{\psiind{A}{i}{-}}

\def\pbaip{\psibind{A}{i}{+}}
\def\pbajp{\psibind{A}{j}{+}}
\def\pbaim{\psibind{A}{i}{-}}
\def\pbajm{\psibind{A}{j}{-}}
\newcommand\fai[4]{{#1}^{{#2}}_{{#3}} ({#4}) }

\def\rilmxz{\fai{\rho}{i l}{-}{x,z}}
\def\rljmxy{\fai{\rho}{l j}{-}{x,y}}
\def\rljmzy{\fai{\rho}{l j}{-}{z,y}}
\def\rijpxy{\fai{\rho}{i j}{+}{x,y}}
\def\rljpxy{\fai{\rho}{l j}{+}{x,y}}
\def\rijmxy{\fai{\rho}{i j}{-}{x,y}}
\def\rljmxy{\fai{\rho}{l j}{-}{x,y}}

\def\aijp{{\alpha^{i j}_+}}
\def\aijm{{\alpha^{i j}_-}}
\def\bijp{{\beta ^{i j}_L}}
\def\bijm{{\beta ^{i j}_R}}

\def\aijmxy{\fai{\aa}{i j}{-}{x,y}}
\def\aijpxy{\fai{\aa}{i j}{+}{x,y}}
\def\ajimyx{\fai{\aa}{j i}{-}{y,x}}
\def\ajipyx{\fai{\aa}{j i}{+}{y,x}}

\def\bijpxy{\fai{\bb}{i j}{L}{x,y}}
\def\bijmxy{\fai{\bb}{i j}{R}{x,y}}
\def\bjipyx{\fai{\bb}{j i}{L}{y,x}}
\def\bjimyx{\fai{\bb}{j i}{R}{y,x}}
\def \bigd{{\cal D}}

\begin{document}
\rightline{NORDITA-1999/35  HE}
\rightline{\hfill August 1999}
\vskip 1.2cm

\centerline{\Large \bf Large N gauge theories and AdS/CFT correspondence}

\vskip 1.2cm

\centerline{\bf Paolo Di Vecchia}
\centerline{\sl NORDITA}
\centerline{\sl Blegdamsvej 17, DK-2100 Copenhagen \O, Denmark}
\vspace{1cm}
\centerline{ {\it Lectures given at the Spring Workshop on Superstring and 
Related Matters, }}
\centerline{\it{Trieste, Italy  (March 1999).}}
\vspace{1cm}
%
%
%
%
%
%

\begin{abstract}
In the first part of these lectures we will review the main aspects of
large $N$ QCD and the explicit results obtained from it. Then, after a
review of the properties of ${\cal{N}}=4$ super Yang-Mills, type IIB string
theory and of $AdS$ space, we briefly discuss the Maldacena conjecture.
Finally in the last part of these lectures we will discuss the finite 
temperature case and we show how "hadronic" quantities as the string tension,
the mass gap and the topological susceptibility can be computed in this 
approach.   
\end{abstract}

\vskip 1cm
%
\sect{Introduction}
\label{intro}
\par
Gravity is described by the Einstein's theory of general relativity, while 
the other interactions are described by  gauge field 
theories. Actually also the theory of general relativity is a gauge theory
corresponding to the gauging of the space-time Poincar{\'{e}} group, while
those that are usually called gauge theories correspond to the
gauging of an internal symmetry. But apart from the fact that they are
both gauge theories does it exist any deeper relation between them? Do they
imply each other in a consistent quantum theory of gravity? 
In the framework of field theory there is no connection; they 
can both exist independently from each 
other, but any field theory involving gravity suffers from the problem of
non-renormalizability. In the framework of string theory, instead, where quantum
gravity makes sense, we see not only that they naturally occur 
together in the same theory, but also that any attempt for constructing a 
string theory without gravity has been until now unsuccessful. This seems to
suggest that the presence of both gravitational and gauge interactions is
perhaps unavoidable in a consistent string theory. 

String theories originated from the attempt of describing the properties of
strong interactions through the construction of the dual resonance model.
It became soon clear, however, that this model in its consistent form, 
that later on was recognized to correspond to the quantization of  a 
relativistic string,
contained all sort of massless particles as gluons, gravitons and others
except a massless pseudoscalar particle corresponding to the pion that in the
chiral limit is
the only massless particle that we expect in strong interactions. 
\par
Actually a model describing $\pi \pi$ scattering in a rather satisfactory way  
was proposed by Lovelace and Shapiro~\cite{LOVE}. According to this
model the three isospin amplitudes for pion-pion scattering are given by:
\[
A^0 = \frac{3}{2} \left[ A(s,t) + A (s,u) \right] - \frac{1}{2} A(t,u)
\]
\beq
A^1 = A(s,t) - A (s,u) \hspace{2cm} A^2 = A (t,u)
\label{isoamp}
\eeq
where
\beq
A(s,t) = \beta \frac{\Gamma (1 - \alpha_s ) 
\Gamma (1 - \alpha_t)}{\Gamma (1 - \alpha_t - \alpha_s )}
\hspace{1cm};\hspace{1cm} \alpha_s = \alpha_0 + \alpha' s
\label{ast}
\eeq
$s = - (k_1 + k_2 )^2 $, $t = - (k_1 - k_3 )^2 $ and $u = - (k_1 - k_4 )^2 $
are the three Mandelstam variables that satisfy the relation: $ s+t+ u = -
\sum_i k_{i}^{2} = 4 m_{\pi}^{2}$. The amplitudes in eq.(\ref{isoamp}) provide
a model for $\pi \pi$ scattering with linearly rising Regge trajectories 
containing three parameters: the intercept of the $\rho$ Regge trajectory 
$\alpha_0$, the Regge slope $\alpha'$ and $\beta$. The first two can be
determined by imposing the Adler's self-consistency condition, that requires
the vanishing of the amplitude when $s=t=u = m_{\pi}^{2}$ and one of the pions
is massless, and  the fact that the Regge trajectory must give the 
spin of the $\rho$ meson that is equal to $1$ when $\sqrt{s}$ is equal to the
mass of the $\rho$ meson $m_{\rho}$. These two conditions determine the Regge
trajectory to be:
\beq
\alpha_s = \frac{1}{2} \left[1 + 
\frac{s - m_{\pi}^{2}}{m_{\rho}^{2} - m_{\pi^{2}}} \right] = 0.48 + 0.885 s
\label{rhotra}
\eeq
Having fixed the parameters of the Regge trajectory the model predicts 
the masses and the couplings of the resonances that decay in $\pi \pi$ 
in terms of a unique parameter $\beta$. The values obtained are in 
reasonable agreement with the experiments. Moreover  one can compute the
$\pi \pi$ scattering lenghts:
\beq
a_0 = 0.395 \beta \hspace{2cm} a_2 = - 0.103 \beta
\label{scatle}
\eeq
and one finds that their
 ratio is within $10 \%$ of the current algebra ratio given by $a_0 /a_2
= - 7/2$. 
The amplitude in eq.(\ref{ast}) has exactly the same form as the four tachyon 
amplitude of the Neveu-Schwarz model with the only apparently minor difference
that $\alpha_0 =1/2$ (for $m_{\pi} =0$) instead of $1$ as in the Neveu-Schwarz 
model. This difference, however, implies that the critical dimension of this
model is $D=4$~\footnote{This can be checked by computing the coupling of the 
spinless particle at the level $\alpha_s =2$ and seeing that it vanishes for
$D=4$.} and not  $D=10$ as in the NS model. In conclusion this model seems to 
be a
perfectly reasonable model for describing low-energy $\pi \pi$ scattering. The
problem is, however, that nobody has been able to generalize it to the 
multipion scattering and to get a string interpretation as instead one has done
in the case of the known string models. 

Because of this and also because of the presence of some features in the 
string model  that are not shared by strong interactions as for
instance the exponential instead of the power decay of the hadronic 
cross-section at large transverse momentum,   
it became clear in the middle of the seventies that string theories could 
not provide a theory for strong
interactions, that in the meantime were successfully described in the
framework of QCD, but could instead be used as a consistent way of unifying all
interactions in a theory containing also quantum gravity~\cite{SCHERK}. 
It turned out  in fact that all five consistent
string theories in ten dimensions all unify in a way or another gravity
with gauge theories. Let us remind shortly how this comes  about. 

The type I theory is a theory of open and closed string.
Open strings have Chan-Paton gauge degrees of freedom located at the 
end points and, because of them, an open string theory contains the usual
gauge theories. On the other hand a pure theory of open strings is not
consistent by itself; non-planar loop corrections generate closed strings
and a closed string theory contains gravity. Therefore in the type I
theory open strings require for consistency  closed strings. This implies that
gravity, that is obtained in the zero slope limit ($\alpha_{cl} ' 
\rightarrow 0$) of closed strings, is a necessary consequence of gauge 
theories, that are obtained in the zero slope limit of the open string theory
 ($\alpha' \rightarrow 0$). Remember that the two slopes are related
through the relation $\alpha_{cl}' = \alpha '/2$.

The heterotic strings is instead a theory of only closed strings that 
contains, however, both supergravity and gauge theories. But in this case 
gravity is the fundamental theory and gauge theories are
obtained from it through a stringy Kaluza-Klein mechanism. The remaining 
consistent theories in ten dimensions are the two type II theories that at the 
perturbative level contain only closed strings and no gauge degrees of freedom. 
However, they also contain non-perturbative objects, the $D$-branes that are 
characterized by the fact that open strings can end on them. Therefore
through the D-branes open strings also appear in type II theories and
with them we get also gauge theories.

In conclusion all string theories contain both gravity and gauge theories
and therefore those two kinds of interactions are intrinsically unified
in string theories. But, since all string theories contain
gravity, it seems impossible to use a string theory to describe strong
interactions. In fact they are described by QCD that does not
contain gravity!!

On the other hand it is known since the middle of the seventies that, if we
consider a non-abelian gauge theory with gauge group $SU(N)$ and we take
the 't Hooft limit where the number of colours $N \rightarrow \infty$,
while the product $g^{2}_{YM} N \equiv \lambda$ is kept fixed~\cite{HOOFT}, 
the gauge theory simplifies because  only planar diagrams
survive in this limit. In the large $N$ limit 
the gauge invariant observables are determined by a master
field~\cite{WITTEN1} that satisfies a classical equation of motion. It has 
also been
conjectured that in this limit QCD is described by a string theory; the
mesons are string excitations that are free when $N \rightarrow \infty$.
This idea is also supported by the experimental fact that hadrons lie on 
linearly rising 
Regge trajectories as required by a string model. The fact that the large
$N$ expansion may be a good approximation also for low values of $N$ as
$N=3$ is suggested by the consistency of its predictions with some 
phenomenological observations as for instance  the validity of the 
Zweig's rule and  the successful  explanation of the $U(1)$-problem in the 
framework of the large $N$ expansion. This is also confirmed by 
recent lattice simulations~\cite{TEPER}. The concrete way in which the 
large $N$
expansion explicitly solves the $U(1)$-problem is reviewed in 
Sect.~\ref{u1pro}. 
The fact, however, that any consistent 
string theories includes necessarily gravity has led to call the
string theory coming out from QCD, the QCD string because, as QCD, it should
not contain gravity. Although many attempts
have been made to construct a QCD string none can be considered sufficiently
satisfactory. This problem has been with us for the last thirty years. In
Sect.~\ref{largeN} we will review the large $N$ expansion in gauge theories 
and the various
arguments that brought people to think that, for large $N$, a string theory
ought to emerge from QCD. Unfortunately, although the large-$N$ expansion 
drastically simplifies the structure of QCD   keeping only the planar diagrams,
it has not yet been possible to carry it out explicitly in the case of 
four-dimensional gauge theories. In order to show some example in which the
large $N$ expansion can be explicitly performed in Sect.~\ref{cpnmo} we 
discuss it in the $CP^{N-1}$ model, where it has allowed us 
to study several important aspects that these models share with QCD as for
instance confinement and the $U(1)$ problem, and in two-dimensional QCD,
where a master field picture emerges and the spectrum of mesons can be
explicitly computed.

Recent studies of D branes have allowed to establish another deep connection
between gravity and gauge theories. In fact, on the one hand, a system of $N$ D
$p$-branes is a classical solution of the low-energy string effective action, 
containing
gravity, dilaton and an antisymmetric R-R $(p+1)$-form  potential. 
The metric corresponding to a D $p$-brane in $D=10$ is given by:
\beq 
(ds)^2 = H^{-1/2} (y) \eta_{\alpha \beta} d x^{\alpha} d x^{\beta} + H^{1/2} (y)
\delta_{ij}d y^{i} d y^{j}
\label{clasol}
\eeq
while the dilaton and RR potential are equal to:
\beq
e^{- (\phi - \phi_0) } = [ H(y) ]^{(p-3)/4} \hspace{1cm}; \hspace{1cm} A_{01 
\dots p}= [H(y)]^{-1}
\label{clasol2}
\eeq
where
\beq
H (y) = 1 + \frac{K_p N}{r^{7-p}} \hspace{2cm} K_p = 
\frac{(2 \pi \sqrt{\alpha '})^{7-p}}{(7-p) \Omega_{8-p}} g_s
\label{acca}
\eeq
with $r^2 \equiv y_i y^i$ and $\Omega_{q} = 2 \pi^{(q+1)/2}/\Gamma[(q+1)/2]$.
The indices $\alpha $ and $\beta$ run along the world volume of the brane, while
the indices $i$ and $j$ run along the directions that are transverse to the 
brane.

On the other hand the low-energy dynamics of a system of $N$ D $p$-branes
is described by the non abelian
version of the Born-Infeld action that is a functional of the transverse 
coordinates of the brane $x^i$ and of a gauge field $A^{\alpha}$ living on 
the brane.
Its complete form is not yet known, but for our considerations we can take it
of the form suggested in Ref.~\cite{TSE}:
\beq
S_{BI} = - \tau_{p}^{(0)} \int d^{p+1} \xi \,\, e^{-\phi} STr \,\, \sqrt{- \det 
\left[ G_{\alpha \beta } + B_{\alpha \beta} + 2 \pi \alpha ' F_{\alpha \beta} 
\right]}
\label{bi3}
\eeq
The brane tension is given by:
\beq
\tau_p \equiv \frac{\tau_{p}^{(0)}}{g_s} = 
\frac{(2 \pi \sqrt{\alpha'})^{1-p}}{2 \pi \alpha' g_s} \hspace{2cm} g_s \equiv
e^{\phi_0}
\label{pten}
\eeq 
where the string coupling constant $g_s$ is identified with the value at 
infinity of the dilaton field. $G_{\alpha \beta}$  and $B_{\alpha \beta} $ are 
the pullbacks of the metric $G_{\mu \nu}$ and of the two-form NS-NS potential 
$B_{\mu \nu}$, while $F_{\alpha \beta}$ is a gauge field living on the brane.
$S Tr $ stands for a symmetrized trace over the group matrices. In addition
to the term given in eq.(\ref{bi3}) the effective action for a D $p$-brane
contains also a Wess-Zumino term that we do not need to consider here. 
By expanding the Born-Infeld action in powers
of $\alpha'$ we find at the second order the kinetic term
for a non abelian gauge field (the $U(N)$ matrices are normalized as 
$Tr (T_i T_j ) = \frac{1}{2} \delta_{ij}$):
\beq
S_{BI} = - \frac{1}{4 g_{YM}^{2}} \int d^{p+1} \xi \,\, F^{a}_{\mu \nu} 
F^{a \mu \nu} \hspace{.5cm};\hspace{.5cm} 
g^{2}_{YM} = 2 g_s (2 \pi)^{p-2} (\alpha ')^{(p-3)/2}
\label{biexp}
\eeq
An interesting property of the D-brane solution in eqs.(\ref{clasol}) and 
(\ref{clasol2}) is that for large values of $r$ the metric  becomes flat.
Therefore, being the curvature small, the classical supergravity description
provides a good approximation of the D brane. 

Based on the previous deep connection between gauge theories and type IIB 
supergravity or more in general type IIB superstring and on the fact that
the metric of a D$3$-brane in the near-horizon limit becomes that of $AdS_5 
\times S^5$ Maldacena~\cite{MALDA} made
the conjecture that actually the low-energy effective action
of a D$3$-brane, that is given by ${\cal{N}}=4$ super Yang-Mills theory in
four dimensions, is equivalent to type IIB string theory compactified on 
$AdS_5 \times S^5$. A detailed discussion of the Maldacena conjecture is
presented in Sect.~\ref{malda}, while Sect.~\ref{AdS} is devoted to general
properties of anti De Sitter space and Sect.~\ref{n=4} to the symmetry
properties of both ${\cal{N}}=4$ super Yang-Mills and type IIB string theory.

The Maldacena conjecture provides for the first time a strong evidence that
a string theory comes out from a gauge theory.
But ${\cal{N}}=4$ super Yang-Mills is in the
Coulomb phase and therefore the emergence of a string has nothing to do with
the confining properties of the theory. In order to get a confining theory we
have to get rid of the conformal invariance of the theory. The simplest way
of doing so is by considering ${\cal{N}}=4$ super Yang-Mills at finite
temperature, i.e. by considering its euclidean version with compactified
time. Since  bosons have periodic and fermions anti-periodic
boundary conditions, in  going to finite temperature, we also break 
supersymmetry. 
In order to deal with ${\cal{N}}=4$ super Yang-Mills at finite temperature
it is necessary to consider a finite temperature version of AdS 
space~\cite{HAWKING}.
This is
what we present in Sect.~\ref{fintemp}, where, following Witten~\cite{WITTEN3},
we actually see that we can identify two manifolds both having as boundary 
the compactified Minkowski four-dimensional space. It turns out that one of
them is dominant at low temperature where the theory is still in the Coulomb 
phase, while the other one is dominant at high temperature where instead the
theory is confining~\cite{WITTEN3}. In the latter case the Wilson loop gives 
a contribution
proportional to the area and from it one can extract a finite string tension.
In this case the theory has a new phase at high temperature characterized by
confinement and by the emergence of a mass gap~\cite{WITTEN3}. From the 
point of view of type
IIB supergravity this is seen as the emergence of another solution of the
supergravity equations, namely the AdS black hole, that becomes dominant at
high temperature, while  empty AdS space is still dominant at low temperature.  
This is presented in sect.~\ref{fintemp} where we also compute the Wilson 
loop,
from which we can extract the string tension, the mass gap and more in
general the discrete spectrum of glue balls. 

In the final section~\ref{4dimym} we discuss a recent proposal by 
Witten~\cite{WITTEN3}
for studying four-dimensional Yang-Mills theory starting from the M-theory
$5$-brane solution and we compute in this approach the topological
susceptibility and the string tension. 

Recent and some of them very detailed reviews on the AdS/CFT conjecture can 
be found in Refs.~\cite{KLEBANOV2,DAEMI,JENS,PAOLO,MALDAETAL}. 
 
\sect{Large $N$ QCD}
\label{largeN}

In this section we discuss some diagrammatical properties of large $N$ 
QCD~\footnote{For a recent review of the large $N$ expansion in QCD see
Ref.~\cite{MANOHAR}.},
we show that, unlike the perturbative expansion, the large $N$ expansion  is 
an expansion according to the topology of the diagrams and not in powers of 
the coupling constant and we see that a picture in terms of an underlying
string theory seems to naturally emerge from it. At the end of this section 
we discuss the emergence of the master field.

QCD is a gauge field theory based on the colour group $SU(3)$.
It is an asymptotically free theory whose coupling constant in perturbation
theory is given by:
\beq
\alpha_s (Q) = \frac{4 \pi}{( \frac{11}{3} N - \frac{2}{3} N_f)\log 
\frac{Q^2}{\Lambda^2}}   
\hspace{1cm} \alpha_s \equiv \frac{g_{YM}^{2}}{4 \pi}
\label{alph}
\eeq
where $N=3$ and $\Lambda  \sim 250\, MeV$ is the fundamental scale of QCD. 
At high energy
($Q^2 >> \Lambda^2$) the coupling constant is small and therefore QCD is well 
described by perturbation theory, but in order to
study its low-energy properties as confinement, chiral symmetry breaking and
the emergence of a mass gap we need non-perturbative methods. One of them 
consists in putting QCD on a lattice and use numerical simulations. In this 
way, however, we get only a numerical  but not a 
concrete understanding of confinement 
based on a definite approximation. Actually, when we formulate gauge theories
on a lattice, it is rather easy to compute various physical quantities in
the strong coupling approximation and it is even not so difficult to compute
several terms of the strong coupling expansion. For instance it is almost 
immediate to show
that the vacuum expectation value of the Wilson loop has a leading term 
proportional to the area of the loop:
\beq
W(I, J) = e^{- IJ \log (N g_{YM}^{2})} \equiv e^{- IJ a^2  \sigma}
\label{Wilsolo}
\eeq 
where we have considered a rectangular Wilson loop with sides of lenghts
$Ia$ and $Ja$ ($a$ is the lattice spacing). According to the Wilson confinement
criterium consisting in the fact that the Wilson loop is proportional to the
area of the loop the behaviour found in eq.(\ref{Wilsolo})
implies that the strong coupling limit of gauge theories confines and
that in this limit the string tension is given by:
\beq
\sigma = \frac{1}{a^2} \log ( N g^{2}_{YM} )
\label{strite}
\eeq
In the same paper~\cite{KW} in which Wilson found that the Wilson loop is 
proportional to the area in strong coupling lattice gauge theory, it was 
also realized that the strong coupling expansion of the Wilson loop can be 
written as a sum over all surfaces as in the relativistic string model. Thus
a string picture emerges from lattice gauge theory for strong coupling. 
However the behaviour of lattice gauge theory for strong coupling has in 
general nothing to do with the continuum limit of the theory, that is
the one we are interested in and that is obtained instead when the lattice
spacing $a$ goes to zero corresponding, because of asymptotic freedom, to a
weak coupling limit:
\beq
a^2 \Lambda_{0}^{2} = e^{- 16 \pi^2/(\beta_0 g^{2}_{YM} (a))} 
\hspace{2cm} \beta_0 = \frac{11}{3} N - \frac{2}{3}N_f
\label{asyfree}
\eeq
where $\Lambda_0$ is the QCD scale in some normalization scheme.
According to the renormalization group a physical quantity as the string
tension should show the same behaviour in terms of the coupling constant
as in eq.(\ref{asyfree}):
\beq
\sigma a^2 = \left(\sigma/\Lambda_{0}^{2} \right) 
e^{- 16 \pi^2/(\beta_0 g^{2}_{YM} (a))}
\label{asyfree4}
\eeq
Monte Carlo numerical simulations have shown~\cite{MCREU} that
confinement is indeed also a property of the   weak coupling limit in which
the continuum theory is supposedly recovered as one can see from the Monte
Carlo data that show the exponential behaviour with the coupling constant
as in eq.(\ref{asyfree4}).

However this limit cannot be performed analytically in some approximation. Up
to now it has only been possible to reach it by numerical simulations.

Since its original formulation~\cite{HOOFT} the
large $N$ expansion has been the most concrete possibility for reaching an
analytical understanding of the non-perturbative aspects of QCD including
its confinement properties. One generates 
a new expansion parameter by introducing $N$ instead of $3$ colours.
This means that we consider an $SU(N)$ instead of an $SU(3)$ gauge theory. 


In order to describe the large $N$ expansion it is convenient to draw QCD 
Feynman
diagrams in an apparently complicated notation~\cite{HOOFT}: a gluon propagator
is drawn as a pair of colour lines (each carrying a label
going from 1 to  $N) $ and a quark propagator as one colour
and one flavour-carrying line. When propagators are joined through
vertices (also written, of course, in double-line notation)
one can count, for each Feynman diagram, its dependence upon
the gauge coupling $g_{YM}$, and the numbers of colour $N $  and of flavours 
$N_{f} $.
In 't Hooft's original expansion one keeps the number of flavours
$N_{f} $, as well as the combination  $g^{2}_{YM} N $, fixed as $N $
goes to infinity. The latter requirement follows from the need
to keep  $\Lambda_{QCD} $  fixed (see eq.(\ref{alph})), ensuring that meson 
masses approach
a finite limit. The requirement of keeping  $N_{f} $ fixed is less
obvious (in Nature, after all, $N <N_{f}) $ but is crucial in
order to have vanishing mesonic widths (they behave like
 $N_{f}/N$) and to establish therefore a connection with tree-level
string theory. Therefore in the following we will keep $N_f$ fixed when
$N \rightarrow \infty$.

By looking at specific examples it is easy to get convinced of the validity
of the following  general properties:
\begin{enumerate}
\item{If one considers a correlation function of gauge invariant operators
and if one looks at its dependence upon $N $  and $g^{2}_{YM}N \equiv 
\lambda  $
one can see that the dependence on $\lambda$ clearly
varies with the order of the diagram, while the  dependence on $N $
 is only sensitive to its topological properties.  Thus the large $N$ 
expansion selects the topology of Feynman diagrams
rather than their order and can pick up, at lowest order, important
non-perturbative effects.}
\item{Non-planar diagrams are down by a factor $1/N^2$ with respect to the
planar ones.}
\item{Diagrams with quark loops are down by a factor $1/N$ with respect to those
without quark loops.}
\end{enumerate}
\par
Because of this the diagrams that dominate in the large $N$ limit are the 
planar ones with the minimum number of quark loops.
\par
Let us consider a matrix element with two gauge invariant operators $J(x)$
involving bilinears of quark fields as for instance ${\bar{\psi}} \psi$ or 
${\bar{\psi}} \gamma_{\mu} \psi$. The dominant connected diagrams contributing
to a correlator
containing two or more $J's$ are the planar ones with only one quark loop filled
in all possible ways by the gluon exchanges. It is easy to see that the two 
and actually also the multipoint correlators are of order $N$ for large $N$
\beq
< J(k) J(-k) >  \sim 0(N)
\label{orden}
\eeq
because the quark loop gives a factor $N$ while all the gluon exchanges
give something that is constant if $\lambda$ is kept fixed. Since the diagram
is planar it is easy to convince oneself that, if one cuts it, the intermediate 
states consist of an ordered set of partons starting from a quark and after 
many gluons ending on an antiquark and that each parton shares colour indices
with his nearest neighbours in the chain.  As a consequence the intermediate
states  are singlets of the gauge group and it is natural to associate them 
with mesons. Intermediate states with two mesons are negligible at large $N$. 
One can then factorize the two-point correlators
in terms of a sum over meson contributions:
\beq
< J(k) J(-k) > = \sum_{n} \frac{a_{n}^{2}}{k^2 - m_{n}^{2}}  \sim 0(N)
\label{orden2}
\eeq
In the perturbative regime we can use perturbation theory where we can 
see that the previous correlator behaves as a logarithm of $k^2$. In order to
reproduce this logarithmic behaviour, we need an infinite number of mesons. In
addition, since the meson masses, being proportional to $\Lambda$, are smooth 
when $N \rightarrow \infty$, i.e.
$m_n \sim 0(1)$, then the meson coupling, corresponding to the probability
amplitude for a current to create a meson, $a_n = <0| J | n> \sim 0(\sqrt{N})$
grows up as $\sqrt{N}$. 
It is natural to associate intermediate states of this kind with
string-like states    of the form
\EQ
|{\cal{ M}}( {\cal{C}}_{xy} ) > \equiv
{\cal{M}} ( {\cal{C}}_{xy})  |0> \equiv {\bar{\psi}} (x) P
 e^{i \int_{{\cal{C}}_{xy}} A_{\mu} dx_{\mu}} \psi (y) |0>
\label{6.1}
\EN
in which the path ${\cal{C}}_{xy}$ can be seen as a string with
quarks at its ends ($P$ denotes a path-ordered exponential and
the trace is performed in group space).

Similarly, for gauge invariant correlation functions of purely gluonic sources,
intermediate states at large $N_c$ have the same colour structure
as              :
\EQ
\label{6.2}
|{\cal{W}}( {\cal{C}})  > \equiv {\cal{W}} ({\cal{C}})| 0> \equiv
  Tr P \exp \left[\oint_{{\cal{C}}} dx_{\mu} A_{\mu}(x) \right]
|0>
\EN
and are thus naturally associated with a closed string described
by the path ${\cal{C}}$.
\par
Let us consider now a correlator involving 3 currents $J$. It will be a function
of the three momenta $p, r, s$ of the three operators. It can contain three 
poles respectively in the variables $p^2, r^2 $ and $s^2$ corresponding to the
masses of the three mesons or only two poles. The terms with three poles 
contains
three couplings $a_n$ and a $3$-meson vertex. Since each $a_n \sim 0 
({\sqrt{N}})$ and the total expression is $0(N)$ then the $3$-meson vertex is
$0(1/\sqrt{N} )$. This means that, for $N \rightarrow \infty$ the mesons are 
an infinite number of stable particles. 
The large $N$ expansion has the nice property of separating the problem of 
the formation
of hadrons connected to quark confinement and the generation of a mass gap
from the problem of their residual interaction. Actually there are also some 
reasons to believe that mesons are excitations of a string. Already the 
representation of a meson given in eq.(\ref{6.1}) is strongly reminiscent of
a string. In addition,  the perturbative expansion in string theory in
terms of the string coupling constant $g_s$ and the large $N$ expansion of gauge
theories in powers of $1/N$ are both topological expansions in the sense that 
they are expansions
according to the topology of respectively the string and the gauge theory 
diagrams. In particular the planar diagrams, that are the dominant ones in
gauge theories, are pretty much reminiscent of the tree diagrams of string 
theory. A tree diagram for the scattering of $M$ mesons in string theory
is of the order $g_{s}^{M-2}$, while the same amplitude for $N \rightarrow
\infty$ in gauge theory is of the order $N^{- 1 -M/2}$. A characteristic 
feature of 
the planar approximation, that we have already seen above, is that the
intermediate states are "irreducible" colour singlet, in the sense
that they cannot be split into two singlets. This is why the
corresponding mesons should have exactly zero width in the large $N$
limit precisely as it is the case in tree-level string theory.
Additional evidence for having a string theory coming out from QCD comes 
from hadron phenomenology. In fact, if mesons are
excitations of strings they will lie on linearly rising Regge trajectories as 
experiments seem to indicate. There are also other aspects of hadron
phenomenology that are well explained (also numerically) by the large-$N$
expansion as for instance the fact that even the heavy hadrons have a small
width relative to their mass. Other ones are
the validity of the Zweig's rule according to which
for instance the meson $\phi$ decays in $k \bar{k}$ rather than in $3$ pions 
as favoured
by the phase space and the numerical explanation of the $U(1)$ problem.
\par
If 't Hooft's considerations can be easily extended from mesons made by a 
quark-antiquark pair
 to glueballs, predicting in particular their existence and narrowness,
                the generalization to baryons is much more subtle.
This is certainly related to the fact that a baryon is, by
definition, a completely antisymmetric object that one can
make out of $N $ quarks. Thus, unlike the mesonic   case, the
baryon's wave-function changes in an essential way with $N $
and one cannot expect the large $N $ limit to be smooth.
Arguments can be given  for the baryon mass to scale indeed
like $N $  and therefore like $1/(1/N). $ If we identify $1/N$ with a
coupling constant, such a behaviour is
reminiscent of the monopole mass, and indeed, Witten (see
Refs.~\cite{WITTEN0,GERSAK}) has
taken up this analogy quite far. Yet, the actual relevance of
large $N $ baryons for the physical nucleon remains to be
proven. 
\par
In conclusion we have seen that the large $N$ expansion 
provides a very natural framework for 
discussing this QCD reinterpretation of the old Dual-String.

In the last part of this section we will briefly discuss the idea of the 
master field.
We have seen above that, if we restrict ourselves to connected diagrams, the
leading term of a correlator involving composites of the type ${\bar{\psi}}
\psi$ is $0(N)$, while that of a correlator involving composites of the gluon is
$0(N^2)$. It is, however, easy to see that disconnected diagrams are in general
dominating. Therefore if we consider the Green's function of a collection of 
Wilson-loop operators as the ones given in eqs.(\ref{6.1}) and (\ref{6.2}):
\EQ
\label{6.5}
<0| {\cal{M}}_{1} \cdots {\cal{M}}_{n} {\cal{W}}_{1} \cdots {\cal{W}}_{m}| 0>
\EN
it is rather simple to prove that the leading large $N$ diagrams
cannot have any propagator joining together two different operators.
As a consequence the VEV in eq. (\ref{6.5}) becomes the product of the
VEV's:
\[
<0| {\cal{M}}_{1} \cdots {\cal{M}}_{n} {\cal{W}}_{1} \cdots {\cal{W}}_{m}| 0>
      =
\]
\EQ
\label{6.6}
=  <0| {\cal{M}}_{1}|0> \cdots <0|{\cal{M}}_{n}|0> <0|{\cal{W}}_{1}|0>
\cdots <0| {\cal{W}}_{m} |0>
\EN
This almost trivial observation actually leads to a very deep
result: the functional integral defining our correlation
functions must be dominated by a \underline{single} field, the
so-called master field~\cite{WITTEN1}.
This result follows immediately
from studying the expectation value of the square of an operator
which can be equal to the square of the expectation value if
and only if the quantum average is completely dominated by
one path (as in the classical theory).

This powerful result gave great hope that the large $N$ limit
of QCD could be solved in closed form. There
is a large literature discussing the many amazing properties and
equations satisfied by the master field. We do not have time to
discuss it further here.

Unfortunately, none of these approaches has lead so far
to an explicit expression  for the large $N$ limit of
four dimensional QCD: nonetheless,
                                    the idea that some kind of string
must come out from QCD is still very popular.
We know, however, that any string model associated with QCD
                                        cannot coincide with the
usual bosonic (or super) string     since these    contain gravity (or
supergravity), i.e. interactions which are not contained in QCD.

These problems have also brought several people to think that the QCD string 
is not infinitely thin but has a
finite cross section. May be the relevant model is some kind of bag
model with stringlike configurations.

Although no string model has yet     been derived in a rigorous way from
QCD, we have presented a number of indications supporting such a
connection. 

In conclusion, the large $N$ expansion is a very promising approach to
understand non-perturbative properties of QCD as confinement, chiral symmetry
breaking and the generation of a mass gap, but, although it drastically 
simplifies the structure of QCD keeping only the planar diagrams, it has not 
been possible to perform it explicitly and arrive to an explicit computation
There are also a number of indications supporting the idea that a string model
is coming out from QCD. This  has been, however, clashing with the fact that
all consistent string models contain gravity, while QCD does not.

\sect{The large $N$ expansion in $CP^{N-1}$ model}
\label{cpnmo}

We have concluded the previous section by seeing that it has not been possible
to explicitly perform the large $N$ expansion in a matrix theory as QCD. We 
call it matrix theory because the gluon field is a matrix of $SU(N)$.  
In the first part of this section we study the properties of a very interesting 
two-dimensional vector model, called the $CP^{N-1}$ model with fermions, 
because,
on the one hand, it has many properties in common with QCD as classical 
conformal  invariance, existence of a 
topological charge, instanton solutions, confinement and $U(1)$ anomaly and, 
on the other hand, it can be explicitly solved in the large $N$ limit.
This model has been very useful in the past as a toy model for QCD.
Although, unlike other two-dimensional models, confinement is in this model a
quantum effect, the study of confinement in this model has not helped very much
to understand confinement in QCD because confinement
in two dimension is substantially different from confinement in four dimensions.
It has instead been  very useful for understanding how to solve the $U(1)$ 
problem in QCD. 

The action of the $CP^{N-1}$ model with fermions is given 
by~\footnote{A review of this model with all relevant references can be found
in Ref.~\cite{PAOLO2}.}:
\beq
S = \int d^2 x \left\{ {\overline{D_{\mu}z}} D_{\mu} z + {\bar{\psi}} \left(
\fslu{D}  - M_B \right) \psi  
- \frac{g}{2 N_F } \left[ ({\bar{\psi}} \lambda^i \psi )^2 + 
({\bar{\psi}} \lambda^i \gamma_5 \psi )^2 \right] \right\}
\label{cpn}
\eeq
The scalar field $z$ has a colour index that transforms according to the 
fundamental representation of $SU(N)$, while the fermion field has a
colour index that takes values from $1$ to $N_F$ together with a flavour index
that transforms according to the fundamental of $SU(N_f )$. $D_{\mu}$ is the 
covariant derivative of a $U(1)$ gauge field $A_{\mu}$ and the scalar field z 
satisfies a constraint:
\beq
 |z|^2 = \frac{N}{2f}  \hspace{2cm} D_{\mu} = \partial_{\mu} + \frac{2ief}{N}
A_{\mu}
\label{covco3}
\eeq 
$e$ is taken to be equal to $1$ in the covariant derivative for $z$.
\par
In order to study the quantum theory of the model one must compute the
generating functional for the euclidean Green's functions given by:
\[
Z(J, {\bar{J}}, \eta, {\bar{\eta}} ) = \int Dz D{\bar{z}} D\psi D{\bar{\psi}}
\delta (|z|^2 - \frac{N}{2f} ) 
\]
\beq
\exp \left\{ -S + \int d^2 x \left[ {\bar{J}} \cdot
z + {\bar{z}} \cdot J + {\bar{\eta}} \cdot \psi  + {\bar{\psi}} \cdot \eta 
\right]\right\}
\label{gen897}
\eeq
One can eliminate the quartic terms for the $\psi$ field by the introduction
of auxiliary fields and one can use the integral form of the $\delta$ 
function.
In this way the action in the previous equation will contain only terms that
are at most quadratic in the fields $z$ and $\psi$. Therefore the functional
integral over those fields can be explicitly performed and one gets:
\[ 
Z(J, {\bar{J}}, \eta, {\bar{\eta}} ) = \int D \alpha D \Phi^i 
D \Phi^{i}_{5} 
\]
\beq
\exp \left\{ - S_{eff} + \int d^2 x \int d^2 y \left[
{\bar{J}}(x)  \Delta_{B}^{-1} (x,y)  J (y) + {\bar{\eta}}(x)  \Delta_{F}^{-1} 
(x,y)  \eta (y)  \right] \right\}
\label{gen8976}
\eeq
where
\beq
\Delta_B = - D_{\mu} D_{\mu} + m^2  - \frac{i}{\sqrt{N}} \alpha 
\hspace{2cm}
\Delta_F = \fslu{D} - M_B -  \frac{\lambda^i}{\sqrt{N_F}} 
\left[ \Phi^i + \gamma_5 \Phi^{i}_{5} \right]  
\label{con5423}
\eeq
and
\beq
S_{eff} = N Tr \log \Delta_B - N_F Tr \log \Delta_F + \int d^2 x\left[ 
i \frac{\sqrt{N} }{2 f} \alpha + \frac{1}{2 g} \left( \Phi^i \Phi^i 
+ \Phi^{i}_{5} \Phi^{i}_{5} \right) \right]
\label{seff}
\eeq
Although the effective Lagrangian is more complicated than the original 
microscopic Lagrangian, it has, however, the advantage of containing directly 
the meson fields $\Phi$ and $\Phi_5$, while the "quark" $\psi$ and the "gluon" 
$z$ fields have been integrated out. The integral over the remaining composite
fiels cannot be performed exactly and we must develop the action for large 
$N$ and  $N_F$ keeping the number of flavours $N_f$ fixed. In addition we must
also make this
expansion around a minimum and such a minimum occurs for a non zero vacuum
expectation value for the fields $\Phi$ and $\alpha$. Actually in the first eq. 
in (\ref{con5423}) the term $m^2$ corresponding to the v.e.v. of $\alpha$ 
has been already explicitly extracted. In the case in which the quark mass
matrix is diagonal we can take the vacuum expectation values as follows:
\beq
< \Phi^0 > = M_s \sqrt{N_F N_f} \hspace{1.5cm} < \Phi^i > = < \Phi^{0}_{5} > =
< \Phi^{i}_{5} >= 0 \hspace{.5cm}, \hspace{.5cm} i \neq 0
\label{vev}
\eeq
The index zero corresponds to the singlet field. Expanding then the effective 
action around such a minimum we get that the leading term for large $N$ is 
equal to:
\[
S^{(1)} = \sqrt{N} i {\tilde{\alpha}} (0) \left[ \frac{1}{2f} - \int 
\frac{d^2 q}{(2 \pi)^2} \frac{1}{q^2 + m^2} \right] + 
\]
\beq
+ 2 {\tilde{\Phi}}^{0}
\sqrt{N_F N_f} \left[ \frac{M_s}{2 g}- M \int \frac{d^2 q}{(2 \pi)^2} 
\frac{1}{q^2 + M^2} \right]
\label{s1}
\eeq
where the tilde indicates the Fourier transform
\beq
{\tilde{\alpha}} = \int d^2 x e^{- i p \cdot x} \alpha (x) \hspace{1cm}
M = M_B + M_s
\label{fouri}
\eeq
The integrals appearing in eq.(\ref{s1}) are ultraviolet divergent. They can be 
regularized by the introduction of a Pauli-Villars cut-off $\Lambda$. 
Then the saddle point condition $S^{(1)} =0$ requires the bare coupling 
constants $f$ and $g$ to vary with $\Lambda$ according to the equations:
\beq
\frac{2 \pi}{f(\Lambda)} = \log \frac{\Lambda^2}{m^2} \hspace{1cm}; \hspace{1cm}
\frac{2 \pi}{g (\Lambda)} = \frac{M_S}{M} \log \frac{\Lambda^2}{M^2}
\label{asy45}
\eeq
which are typical of an asymptotic free theory. In addition, as also in QCD,
there is a dimensional transmutation because in the quantum theory the
dimensionless coupling constants $f$ and $g$ are traded with the two masses
$m$ and $M$.

Having eliminated the term $S^{(1)}$ we can consider the quadratic part of
$S_{eff}$ that is independent of $N$ and $N_F$ and that is given by:
\[
S^{(2)} = \frac{1}{2} \int d^2 x \int d^2 y \left\{  \alpha (x) 
\Gamma^{\alpha} (x-y)
\alpha (y) +  A_{\mu} \Gamma^{A}_{\mu \nu} (x-y) A_{\nu} (y) + \right.
\]
\beq
\left. \Phi^{i} \Gamma^{\Phi}_{ij} (x-y) \Phi^{i} (y) +
\Phi^{i}_{5} \Gamma^{\Phi_5}_{ij} (x-y) \Phi^{i}_{5} (y) + 2 A_{\mu} (x)
\Gamma^{A \Phi}_{\mu} \Phi^{0}_{5} \right\}
\label{s287}
\eeq
where the Fourier transforms of the inverse propagators  are given by:
\beq
{\tilde{\Gamma}}^{\alpha} = A (p;m^2 ) =\frac{1}{2 \pi \sqrt{p^2 (p^2 + 4 m^2)}}
\log \frac{\sqrt{p^2 + 4 m^2} + \sqrt{p^2}}{\sqrt{p^2 + 4 m^2} - \sqrt{p^2}}
\label{gaal}
\eeq
\beq
{\tilde{\Gamma}}^{A}_{\mu \nu} = \left(\delta_{\mu \nu} - 
\frac{p_{\mu} p_{\nu}}{p^2} \right) \left[ (p^2 + 4 m^2) A(p;m^2) - 
\frac{1}{\pi} - \frac{N_F N_f}{N} e^2 \left( 4 M^2 A(p;M^2 ) - \frac{1}{\pi}
\right) \right]
\label{gaa}
\eeq
\beq
{\tilde{\Gamma}}^{\Phi}_{ij} = \delta_{ij} \left[ \epsilon + ( p^2 + 4 M^2)
A(p;M^2) \right]
\label{gaphi}
\eeq
\beq
{\tilde{\Gamma}}^{\Phi_5}_{ij} = \delta_{ij} \left[ \epsilon + p^2  A(p;M^2) 
\right]
\label{gaphi5}
\eeq
and
\beq
{\tilde{\Gamma}}^{A \Phi}_{\mu} = - 2 \epsilon_{\mu \nu} p_{\nu} e M 
\sqrt{\frac{N_f N_F }{N}} A(p;M^2) 
\label{gaphia}
\eeq
In conclusion the leading term in  $1/N$ describes a free theory of 
"mesons" that are composite fields of the fundamental "quark" and "gluon" 
fields.
Higher order terms in the large $N$ expansion will 
describe the meson interaction.
\par
Let us now discuss the physical properties of this model. Because of
asymptotic freedom its short distance properties are completely analogous
to those of QCD. Unlike QCD we are able in this case to analytically study
for large $N$ and $N_F$ also its low energy properties. In particular by
using the low-energy expansion:
\beq
A(p;m^2) \sim \frac{1}{4 \pi m^2} \left[ 1 - \frac{2}{3} \frac{p^2}{4 m^2} 
+ \dots \right]
\label{loenex}
\eeq
we can extract from eq.(\ref{s287}) the low energy effective Lagrangian that
in the simplified case where $N_F = e=1$ is given by:
\[
L_{eff} = \frac{1}{2} \left[ (\partial_{\mu} \Pi^{i} )^2 + m_{\pi}^{2} 
(\Pi^i )^2 \right] + \frac{1}{2} 
\left[ (\partial_{\mu} \sigma^{i} )^2 + (m_{\pi}^{2}+
4 M^2)  (\sigma^i )^2 \right] +
\]
\beq
+\frac{1}{8 \pi m^2} \alpha^2 + \frac{1}{24 \pi m^2} F^2 + i 
\sqrt{\frac{2 N_f}{N}} F_{\pi} F \cdot S
\label{effe453}
\eeq   
where
\beq
F = \epsilon_{\mu \nu} \partial_{\mu} A_{\nu} \hspace{1cm}
\Pi^i = \frac{1}{2 \sqrt{\pi} M} \Phi^{i}_{5} \hspace{1cm}
\Pi^0 \equiv S \hspace{1cm} \sigma^{i} = \frac{1}{2 \sqrt{\pi}M} \Phi^i
\label{defi792}
\eeq
and 
\beq
F_{\pi} = \frac{1}{\sqrt{2 \pi}} \hspace{2cm}m_{\pi}^{2} = 4 \pi \epsilon M^2
\label{fpi7}
\eeq
An important property of this model is the generation of a kinetic term for
the vector field $A_{\mu}$ that was not present in the classical theory. In
two dimension this implies the generation of a confining linear potential with
string tension equal to $\sigma = (12 m^2 \pi)/N$. The factor $1/N$ comes from
the coupling between "coloured" states and $A_{\mu}$. Another important property
that is pretty much related to the previous one is the appearence of a
dependence on the $\theta$ vacuum parameter in the large $N$ expansion. 
In fact if we introduce the topological charge density
\beq
q(x) = \frac{1}{2 \pi \sqrt{N}} F (x)
\label{topcha3}
\eeq
where $F$ is the field defined in eq.(\ref{defi792}), we neglect 
all terms in eq.(\ref{effe453}) that include mesonic fields and we add
a term with the vacuum $\theta$ parameter we are led to
consider the following effective Lagrangian:
\beq
L_{eff} = \frac{1}{2} \frac{\pi N}{3 m^2} q^2 + i \theta q + q J
\label{theta3}
\eeq
where $J$ is an external source. The algebraic equation of motion for $q$
that one gets from the previous Lagrangian is
\beq
q= - \frac{3 m^2}{\pi N} \left( J + i \theta \right)
\label{qso}
\eeq
Inserting it in eq.(\ref{theta3}) we get the following generating functional:
\beq
Z (J, \theta) \equiv e^{- W (J, \theta)} = e^{\frac{3m^2}{2\pi N} \int d^2 x
(J + i \theta)^2}
\label{zetaw}
\eeq
From it putting $J=0$ we can extract the vacuum energy 
\beq
E (\theta) \equiv W(\theta , 0 ) = \frac{3 m^2}{2 \pi N} \theta^2
\label{etheta}
\eeq
the one-point function for the topological charge density
\beq 
< q (x) >_{\theta} = i \frac{3 m^2}{\pi N} \theta
\label{qtheta}
\eeq
and the two-point function for $q$
\beq
< q(x) q(y) > = \frac{3 m^2}{\pi N} \delta (x-y)
\label{qq3}
\eeq
Notice that the vacuum energy has the form 
\beq
E (\theta ) = N F( \theta/N ) \hspace{2cm} F (x) = x^2
\label{theta/N}
\eeq
where the factor $N$ in front counts just the number of degrees of freedom.
From eq.(\ref{qq3}) we can compute the topological susceptibility:
\beq
< q(x) \int d^2 y  q(y) > = \frac{d^2 E (\theta)}{ d \theta^2} =
\frac{3 m^2}{\pi N}
\label{toqq}
\eeq
Another important property
of this model is the presence in the effective low-energy Lagrangian in 
eq.(\ref{effe453}) of a mixed term with the singlet axial field and
the vector field that is fundamental for the resolution of the $U(1)$ problem.
We do not discuss further this here since in the next section we will be 
discussing its resolution in QCD.

In the second part of this section we consider QCD in two dimensions $(QCD_2)$
and we show that in the light cone gauge it is possible to 
reformulate it completely in terms of a bilocal mesonic 
field~\cite{GUTI,CAVICCHI}. We then show that the master field, corresponding 
to the vacuum expectation value of 
the bilocal mesonic field, is fixed in the limit of a large number of 
colours by a saddle point equation whose solution is  equal to the fermion 
propagator constructed in the original paper by 't Hooft~\cite{tH}.
Considering then the quadratic term containing the fluctuation around the 
saddle point it is possible to show that the equation of motion constructed 
from it gives exactly the integral equation found in Ref.~\cite{tH} for the 
mesonic spectrum.     

We consider the action
\beq
S=\int d^2x~
\left\{ -{1\over 4g^2_0}tr(F^{\mu\nu}F_{\mu\nu})
+\pb^i(i\fslu{D}  - m^i  )\pp^i \right\}
\lbl{s-prima} 
\eeq
where $F_{\mu\nu}=\part_\mu A_\nu-\part_\nu A_\mu +i[A_\mu,A_\nu]$,
$i\fslu{D}_{A B}= \gamma_{\mu} (i\part_\mu \uno_{A B}-A^a_\mu T^a_{A B})$,
$a,b=1...N^2-1$ are the indices of the adjoint representation of the
colour group, $A,B=1...N $ run over the fermionic representation 
of the colour $SU(N)$.
If we choose the gauge $A^a_-=0$ and we normalize the trace over the 
fundamental representation to one, we can rewrite the previous 
action\footnote{{\bf Conventions.}
$$
x^\pm=x_\mp=\usrd(x^0\pm x^1)~~~~
A^\mu B_\mu=
A_0 B_0 - A_1 B_1=
A_+ B_- + A_- B_+
$$
$$
\gamma_+=\gamma^-=\mat{0}{\rd}{0}{0}~~~~
\gamma_-=\gamma^+=\mat{0}{0}{\rd}{0}~~~~
\gamma_5=-\gamma_0\gamma_1=\mat{1}{0}{0}{-1}~~~~
P_{R,L}={1\pm\gamma_5\over2}
$$
$$
\pp=\vett{\pp_+}{\pp_-}~~~~
\pb=\vet{\pb_-}{\pb_+}~~~~
$$
$$
\cc\pb=-\usrd
\mat
{\rd\pb P_R\cc}
{\pb \gamma_-\cc}
{\pb \gamma_+\cc}
{\rd\pb P_L\cc}
$$
}as
\begin{eqnarray}
S=\int&d^2x&
\biggl\{
{1\over 2g^2_0}(\dsi A^a_+)^2
+i\rd (\pbaip\dta\paip +\pbaim\dsi\paim)
\nonumber\\
&-&
m^i \pbaim\paip -
m^i \pbaip\paim
-A^a_+~\rd\pbaip T^a_{A B}\pbip
\biggr\}
\lbl{s-gauge}
\end{eqnarray}
Integrating over $A^a_+$ we get 
\begin{eqnarray}
S&=&
\int d^2x~
\biggl\{
i\rd (\pbaip\dta\paip +\pbaim\dsi\paim)
- m^i \left( \pbaim\paip + \pbaip\paim \right)
\biggr\}
\nonumber\\
&&-g^2_0
\int d^2x~d^2y~
G(x-y)~
\pbaip(x) T^a_{A B}\pbip(x)~
\psibind{C}{j}{+}(y) T^a_{C D}\psiind{D}{j}{+}(y)=
\nonumber\\
&=&
\int d^2x~
\biggl\{
i\rd (\pbaip\dta\paip +\pbaim\dsi\paim)
-m^i \left( \pbaim\paip + \pbaip\paim \right)
\biggr\}
\nonumber\\
&+& g^2_0 x_R
\int d^2x~d^2y~
G(x-y)~
\biggl\{
\pbaip(x)\pajp(y)~
\psibind{B}{j}{+}(y) \pbip(x)
\nonumber\\&&~~~~
-{R\over N_R}\pbaip(x)\paip(x)~
\psibind{B}{j}{+}(y) \pbjp(y)
\biggr\}
\lbl{s}
\end{eqnarray}
where we used $\sum_a T^a_{A B} T^a_{C D}=
x_R(\dd_{B C} \dd_{A D}+{R\over N}\dd_{A B} \dd_{C D})$ valid for the 
fundamental representation and
where $G(x)=-{1\over 2}\dd(x^+)\modu{x^-}=\ft{2}{k}{x}{1\over k_-^2}$.

The interaction term suggests to introduce the composite field
\beq
\rijmxy
=\sum_A \pbaj(y)~{\gamma_{-}}~\pai(x)
\lbl{current1}
\eeq
and its partners
\begin{eqnarray}
\rijpxy
=\sum_A \pbaj(y)~{\gamma_+}~\pai(x)
\nonumber\\
\sigma^{i j}(x,y)
=\sum_A \pbaj(y){\uno}~\pai(x)
\nonumber\\
\sigma_5^{i j}(x,y)
=\sum_A \pbaj(y)~{\gamma_5}~\pai(x)
\nonumber\\
\sigma_{R,L}^{i j}(x,y)
= { {\sigma^{i j}(x,y) \pm \sigma_5^{i j}(x,y)}\over \rd}
\lbl{current2}
\end{eqnarray}
Now we want to change variables in the functional integral and integrate 
over the mesonic fields $\rho$ and $\sigma$ instead of the original quark
field $\psi$. We need to compute the jacobian of the transformation 
from the $\pb,\pp$ to the $\rr,\sigma$ that is given by
\[
J[\rhp,\rhm,\sgp,\sgm]=
\int[\bigd\pbaip ~\bigd\paip~\bigd\pbaim~\bigd\paim]
\]
\[
\prod_{^{x y}_{i j}}
\dd[\rijmxy-\rd\sum_A \pbajp(y)\paip(x)]~
\prod_{^{x y}_{i j}}
\dd[\rijpxy-\rd\sum_A \pbajm(y)\paim(x)]
\]
\[
\prod_{^{x y}_{i j}}
\dd[\sigma^{ij}_{R} (x,y) -\rd\sum_A \pbajm(y)\paip(x)]~
\prod_{^{x y}_{i j}}
\dd[\sigma^{ij}_{L} (x,y) -\rd\sum_A \pbajp(y)\paim(x)]
\]
\[
=\int[\bigd\pbaip ~\bigd\paip~\bigd\pbaim~\bigd\paim]
[\bigd\aijp~\bigd\aijm~\bigd\bijp~\bigd\bijm]
\]
\[
e^{
\ajimyx[\rijmxy-\rd\sum_A \pbajp(y)\paip(x)]~
+~\ajipyx[\rijpxy-\rd\sum_A \pbajm(y)\paim(x)]
}
\]
\beq
e^{
\bjipyx[\sigma^{ij}_{R}(x,y)-\rd\sum_A \pbajm(y)\paip(x)]~
+~\bjimyx[\sigma^{ij}_{L} (x, y) -\rd\sum_A \pbajp(y)\paim(x)]}
\lbl{jacobian}
\eeq
where the sum over the flavour and space time indices is understood.

If we introduce the the matrices
\beq
\begin{array}{c}
M=||M_{P Q} ||=\mat{\bijpxy}{\aijpxy}{\aijmxy}{\bijmxy}
\\
U=||U_{P Q} ||=\mat{\sigma^{ij}_{R} (x, y) }{\rijmxy}{\rijpxy}{\sigma^{ij}_{L}
(x, y)}
\\
  \begin{array}{cc}
  \PB^A_Q=\vet{\pbajm(y)}{\pbajp(y)} &
  \PB^A_P=\vett{\paip(x)}{\paim(x)}
  \end{array}
\end{array}
\eeq
where $P\equiv(x i \aa)$ and $Q\equiv(y j \bb)$,
we can rewrite the exponent of the integrand in eq.(\ref{jacobian}) as
\begin{eqnarray}
J[U]&=&\int [d \PB^A d\PP^A][dM]
\exp{[
Tr(MU)
-\rd\PB^A M\PP^A
]}
\nonumber\\
&\propto&\int [d M]
\exp{[
Tr(MU)
+N Tr \log M
]}
\end{eqnarray}
where $N$ is the dimension of the fermionic representation and
$Tr\equiv tr_x~tr_i~tr_\aa$. 
Evaluating this integral with the saddle point method we get
\beq
J[U]\propto\exp[-N Tr\log U]
\eeq
where we have neglected non leading contributions in $N$.

If we define the matrix
\beq
D=||D_{P Q}|| =
\mat
{- m^{ij}\dxy}
{i~\dd^{i j}~\part_{x^-}\dxy}
{i~\dd^{i j}~\part_{x^+}\dxy}
{- m^{ij} \dxy}
\label{def-d}
\eeq
where $m^{ij} \equiv \frac{1}{\sqrt{2}}m^i \delta^{ij}$,
and we rescale the master field $U\rightarrow N U$,  we can rewrite 
the effective action as
\[
{1\over N}S_{eff} =
Tr(D U+i \log U)
+{1\over 2}g^2
\int d^2x~d^2y~G(x-y)~
U_{(x i 1),(y j 2)}~U_{(y j 1),(x i 2)}
\]
\beq
-{1\over 2N}g^2R
\int d^2x~d^2y~G(x-y)~
U_{(x i 1),(x i 2)}~U_{(y j 1),(y j 2)}
\label{eff-s}
\eeq
where $g^2=g^2_0 x_R N$.

Varying the effective action with respect  to $U_{Q P}$, we get the 
equation for the master field, that in the leading order in $N$ is equal to
\beq
D_{P Q}+i (U^{-1})_{P Q}
+g^2~\dd_{\aa,2}\dd_{\bb,1}
G(x-y)~
U_{(x i 2),(y j 1)}
=0
\label{eq231}
\eeq
Multiplying it with $U$, we get immediately 
\beq
D U_{P Q}+i \uno_{P Q}
+g^2~\dd_{\aa,2}
\int d^2z~
~G(x-z)
~U_{(x i 1),(z k 2)}
~U_{(z k 1), Q}
=0
\label{equ654}
\eeq
Writing explicitly these equations we find
\beq
  i\part_{x^+}\rijmxy
  -m^{i l}\sigma^{lj}_{L} (x, y)
  +g^2\int d^2z~G(x-z)\rilmxz\rljmzy +i~\dd^{i j}~\dxy =0
\label{eqn542}
\eeq
\beq  
i\part_{x^-}\sigma^{ij}_{L} (x,y)
  -m^{i l}\rljmxy
  =0
\label{eqn532}
\eeq
\beq
   i\part_{x^-}\rijpxy
  -m^{i l}\sigma^{lj}_{R} (x,y)
  +i~\dd^{i j}~\dxy
  =0
\label{eqn497}
\eeq 
\beq
  i\part_{x^+}\sigma^{ij}_{R}(x,y)
  -m^{i l}\rljpxy
  +g^2\int d^2z~G(x-z)\rilmxz\sigma^{lj}_{R} (z, y)
  =0
\lbl{propagatori}
\eeq
In particular if we eliminate $\sigma^{ij}_{L} (x,y)$ from the first equation 
using the second one, we get the 
fundamental equation
\[
i\part_{x^+} \rijmxy
  +i(m\cdot m)^{i l}~\int~d^2z~\dd(x^+-z^+)~\theta (x^--z^-)~\rljmzy
\]
\beq
  +g^2\int d^2z~G(x-z)\rilmxz\rljmzy
  +i~\dd^{i j}~\dxy
=0
\lbl{prop}
\eeq
In order to solve this equation it is better to pass to momentum space.
Since $\rijmxy=\rd <0|\sum_A \pbajp(y)\paip(x)|0>$ and the vacuum is 
translationally invariant, we need only one momentum for the Fourier 
transform of $\rijmxy$.
The previous equation (\ref{prop}) becomes
\beq
\left[
-p_+ \dd^{i k} +{(m \cdot m)^{i k}\over p_-}
+g^2\int d k~G(k)~\rr^{i k}_-(p-k)
\right]
\rr^{k j}_-(p)
+i \dij
=0
\lbl{prop2}
\eeq
and it suggests to set 
\[
\rijmxy =
\ft{2}{p}{(x-y)}\rr^{i j}_-(p)=
\]
\beq
= \dij \ft{2}{p}{(x-y)} 
{2i~p_-\over 2p_+ p_- -2(m \cdot m)^i -p_- \GG(p)+i\ee}
\lbl{def-mom-r}
\eeq
With this substitution eq. (\ref{prop2}) becomes eq. (\ref{def-d}) of 
ref. \cite{tH}:
\beq
\GG(p)= {4g^2\over(2\pi)^2}\int {d^2k\over k_-^2} 
{i(p_- + k_-)\over 2(p+k)_+(p+k)_- -2(m \cdot  m)^i -(p+k)_- \GG(p+k)
+i\ee}
\lbl{def-GG}
\eeq
The explicit solution yields
\beq
\GG(p)=\GG(p_-) = {g^2\over \pi} 
\left( {sgn(p_-)\over \lambda} - {1\over p_-} \right)
\label{ggp6}
\eeq
where $\lambda$ is an infra-red cutoff introduced in ref. \cite{tH}.

Inserting eq. (\ref{def-mom-r}) in eqs.(\ref{eqn532}), (\ref{eqn497}) and 
(\ref{propagatori}), we get the Fourier transform of the master field
\beq
U^{i j}_0(p)=
{i~\dd^{i j}\over 2p_+ p_- -2(m \cdot m)^i -p_- \GG(p)+i\ee}
\mat{-2 m^{i}}{2p_-}{2p_+ -\GG(p)}{-2 m^{i}}
\eeq 
where $\Gamma (p)$ given in eq. (\ref{ggp6}). $U_0 (p)$ is the master field of 
QCD$_2$ that  is 
identified with the vacuum expectation value of the quark propagator.

Let us now consider  the mass spectrum of the theory, i.e. the fluctuations 
around the master field.
To this purpose we write $U=U_0+{1\over\sqrt{N }}\dd U$, and we 
consider the terms in the effective action  that are $O(1)$ in $N$.
They are given by the quadratic terms in the fluctuation $\dd U$:
\[
S_{eff}^{(2)} =
-{i\over2}Tr(U_0^{-1}\dd U U_0^{-1}\dd U)
+{1\over 2}g^2
\int d^2x~d^2y~G(x-y)
~\dd U_{(x i 1),(y j 2)}
~\dd U_{(y j 1),(x i 2)}
\]
\beq
-{g^2 R\over 2}\int d^2x~d^2y~G(x-y)U_{0(xi1),(xi1)}U_{0(y j2),(y j2)}
\label{eqn610}
\eeq
The last term in the previous equation does not depend on $\delta U$
and therefore will be neglected.
The spectrum of the theory is determined by the equation of motion for 
the field $\dd U$ that is given by
\beq
i~\dd U_{\aa\bb}^{i j}(x,y)
=
 g^2
\int d^2u~d^2v~
U_{0~~\aa,2}^{i k}(x-u)~
G(u-v)~\dd U_{1 2}^{k l}(u,v)~
U_{0~~1, \bb}^{l j}(v-y)
\label{eqn653}
\eeq
and that in Fourier space leads to\footnote{
We define
$$\dd U(x,y)=\int {d^2 r\over (2\pi)^2}{d^2 s\over (2\pi)^2}
e^{ir\left({x+y\over2}\right)+is\left(x-y\right)}
\dd{\tilde U}(r,s)
$$
In the following we suppress the tilde over the Fourier transformed 
fields.
}
\beq
\dd U^{i j}_{\aa\bb}(r,s) = 
-i~g^2{T^{(i j)}_{\aa\bb} (s_-+{r_-\over 2}, s_--{r_-\over 2}) 
\over \Delta^{i}(s+{r\over 2}) \Delta^{j}(s-{r\over 2}) } 
\int {d^2k\over (2\pi)^2} {1\over k_-^2 }~\dd U^{i j}_{1 2}(r,s-k)
\label{eqn748}
\eeq
(no sum over $ i $ and $ j$ ), where
\beq
\Delta^i(p) = 2p_+p_- -2(m \cdot m)^{i} - p_-\GG(p) + i\epsilon
\label{eqn741}
\eeq
 and
\beq
T^{(i j)}_{\aa\bb}(p_-,q_-) = \mat{4p_-m^{j}}{-4p_-q_-}
                                  {-4m^{i}m^{j}}{4m^{i}q_-}
\label{eqn652}
\eeq

Following 't Hooft \cite{tH}, we integrate  both sides of eq.(\ref{eqn748})
over the variable $s_+$ and 
 defining the gauge invariant field\footnote{
Notice that this is equivalent to set $x^+=y^+$ in $\dd U(x,y)$, 
thus obtaining a gauge invariant object. If $x^+\ne y^+$ then $U(x,y)$
is not gauge invariant under the residual gauge transformations.
}
\beq
\vf^{i j}_{\aa\bb}(r,s_-) = \int {d s_+\over 2\pi}~ \dd U^{i j}_{\aa\bb}(r,s) 
\label{eqn6505}
\eeq
we get choosing $r_->0$
\[
\vf^{i j}_{\aa\bb}(r,s_-) = 
g^2
{ T^{(i j)}_{\aa\bb} (s_-+{r_-\over 2}, s_--{r_-\over 2}) 
  \over 4|s_-+{r_-\over 2}||s_--{r_-\over 2}| } 
\left[ { M_i^2\over 2|s_-+{r_-\over 2}|} 
     + { M_j^2\over 2|s_--{r_-\over 2}|}
     +{g^2\over\pi\lambda} - r_+ \right]^{-1}
\]
\beq
\theta(s_-+{r_-\over 2}) \theta({r_-\over 2}- s_-)
\int {dk_-\over 2\pi k_-^2}~\vf^{i j}_{1 2}(r,s_--k_-)
\label{equt}
\eeq
where
\beq
M^2_i=2(m \cdot m)^i-{g^2\over\pi}
\label{ers}
\eeq
In the sector $(\aa,\bb)=(2,1)$ it yields the 't Hooft equation 
(eq. (15) of ref. \cite{tH}) when one identifies the Fourier transform of 
$\rijmxy$ with $\pp(p,r)$.
In the other sectors requiring the cancellation of the IR cutoff $\lambda$, we
get
\beq
\vf^{i j}_{\aa\bb}(r,s_-) = 
{T^{(i j)}_{\aa\bb} (s_-+{r_-\over 2}, s_--{r_-\over 2}) 
\over 4|s_-+{r_-\over 2}||s_--{r_-\over 2}| }
~\vf^{i j}_{1 2}(r,s_-)  
\label{util}
\eeq
Performing the same straightforward manipulations as in ref. \cite{tH}, 
one is led to an
integral equation for the mass spectrum ( $\vf = \vf_{1 2} $; 
we rescale $ s_- = r_-(x-{1\over2}) $ and define $\mu^2 = 2r_+r_-$ ):
\beq
\mu^2 \vf^{i j}(x) = 
\left[{M_i^2\over x} +  {M_j^2\over (1-x)}\right]\vf^{i j}(x)
-{g^2\over\pi}~P\int_0^1 {\vf^{i j}(y)\over (y-x)^2}d y
\lbl{tH-eq}
\eeq
that is the famous 't Hooft equation, with a discrete spectrum of eigenvalues
labelled by an integer $n$ such that 
$\mu^2_{n} \approx g^2\pi~n,~for~ n\to \infty$.
 
In the other sectors we get the same equation for the mass spectrum, but
the mesonic fields change according to

\beq
\vf_{\aa\bb}^{i j}(x) = {\cal C}^{(i j)}_{\aa\bb}(x)\vf^{i j}(x)
\label{poi}
\eeq
with
\beq
{\cal C}_{\aa\bb}^{(i j)} (x) = 
\mat{{m^{j}_R \over (1-x) r_-}}{1}
    {-{m_L^{i}m_R^{j}\over x(1-x) r_-^2}}{-{m^{i}_L \over x r_-}}
\label{fineca}
\eeq

In conclusion in this section we have considered two two-dimensional models, 
the $CP^{N-1}$
model and $QCD_2$, in which the large-$N$ expansion can be explicitly done,
and we have studied their properties for $N \rightarrow \infty$.
In particular in the case of $QCD_2$ we have constructed the master field and
the spectrum of mesons in the large-$N$ limit.

\sect{$U(1)$ problem}
\label{u1pro}

In this section we discuss the resolution of the $U(1)$ problem in the 
framework of the large $N$ expansion of QCD.
In addition to colour gauge symmetry  QCD has also a flavour symmetry. 
If the quark mass matrix is zero QCD is invariant
under the transformations corresponding to independent $U(N_f )$ rotations of 
the right and left parts of the quark field:
\beq
\psi_L \equiv \frac{1 - \gamma_5}{2} \psi \rightarrow U_L \psi_L \hspace{1cm} 
\psi_R \equiv \frac{1 + \gamma_5}{2} \psi \rightarrow U_R \psi_R 
\label{rota34}
\eeq
where both $U_R$ and $U_L$ are $U(N_f )$ matrices. This $U_L (N_f ) \otimes 
U_R (N_f )$
symmetry of the QCD action is called chiral symmetry. In the quantum theory
QCD has an anomaly given by~\footnote{The fact that the resolution of the
$U(1)$ problem is intimately connected to the existence of the axial anomaly
was suggested in Ref.~\cite{THOOFT3}.}:
\beq
\partial_{\mu} J_{5}^{\mu} = 2 N_f q(x) \hspace{1cm}
q(x) = \frac{g^2}{64 \pi^2} \epsilon^{\mu \nu \rho \sigma} F_{\mu \nu}^{a} 
F_{\rho \sigma}^{a} 
\label{axiano}
\eeq
where $q(x)$ is the topological charge density of QCD. As a consequence, QCD
with massless quarks has only a $SU_L (N_f ) \otimes SU_R (N_f ) \otimes 
U_{V} (1)$ where $U_{V} (1)$ corresponds to the baryonic number conservation.
In the real world chiral symmetry can only be an approximate symmetry because 
quarks have a non 
zero mass. However, if we restrict ourselves to the three light flavours, it 
is an 
approximate symmetry because their masses are small
with respect to the QCD scale $\Lambda_{QCD}$. Since, however, this symmetry 
is not seen in the spectrum (there is no scalar particle that has approximately
the same mass of the pion!), it is assumed (and this assumption is 
confirmed by lattice numerical simulations) that in QCD it is spontaneously 
broken. The vectorial $SU(N_f )$ symmetry, that is left unbroken, is instead 
an approximate classification symmetry for the hadrons. 
As a consequence of the spontaneous breaking of chiral symmetry we get that
the pseudoscalar mesons are the quasi Goldstone bosons corresponding to the
spontaneous breaking of chiral symmetry and their low energy interaction can 
be exactly computed. In particular at low energy
we can neglect all hadrons except those that are massless in the
chiral limit. The effective Lagrangian describing the quasi Goldstone bosons 
is that of the non linear $\sigma$-model:
\beq
L = \frac{1}{2} Tr \left( \partial_{\mu} U \partial^{\mu} U^{-1} \right)
+ \frac{F_{\pi}}{2 \sqrt{2}} Tr \left(M U + M^{\dagger} U^{\dagger} \right)
\label{sigma4}
\eeq  
where $U$ is a $3 \times 3$ or $N_f \times N_f $ matrix containing the fields 
of the pseudoscalar mesons:
\beq
U = \frac{F_{\pi}}{\sqrt{2}} e^{i \sqrt{2}\Phi /F_{\pi} }\hspace{1cm}
\Phi = \Pi^i \lambda^i + \uno S/\sqrt{N_f} \hspace{1cm} Tr \left(\lambda^i 
\lambda^j \right) = \delta^{ij}
\label{u786}
\eeq
while the mass matrix $M$ can be taken to be diagonal:
\beq
M_{ij} = \mu_{i}^{2} \delta_{ij}
\label{mass34}
\eeq
In the chiral limit ($M =0$) the Lagrangian in eq.(\ref{sigma4})  is 
invariant under chiral transformations that transform $U \rightarrow U_L U 
U_R$ and therefore cannot be the effective Lagrangian
for low-energy QCD because it does not contain the $U(1)$ anomaly in 
eq.(\ref{axiano}). In order to have the anomaly equation satisfied we must add
to the previous Lagrangian terms that also include the
field corresponding to the topological charge $q(x)$ appearing in the r.h.s. 
of the anomaly equation. On general ground we can write the following 
Lagrangian:
\beq
L = \sum_{i=0}^{\infty} L_{i} (U) [ q(x) ]^{i}
\label{gene45}
\eeq
where the first term with $i=0$ is equal to the kinetic term in 
eq.(\ref{sigma4}).
Neglecting derivative terms, that are irrelevant at low energy, requiring 
parity conservation and imposing that the axial anomaly is reproduced 
(this implies that, under an axial $U(1)$ transformation with angle $\alpha$, 
the previous Lagrangian transforms as $L \rightarrow
L - 2 \alpha N_f  q(x)$) we get that all terms of the sum with even indices are
invariant under the complete $U(N_f ) \otimes U(N_f )$ symmetry, all the
terms with odd indices are vanishing except the first one that is given by
\beq
L_1 = \frac{i}{2} q(x) Tr \left[\log U - \log U^{-1} \right]  
\label{li12}
\eeq
This term precisely reproduces in the effective theory the anomaly equation.
In order to have additional restrictions we need to use the large $N$
expansion. Using the arguments developed in sect.~\ref{largeN} it is easy to 
check the following behaviour with $N$:
\beq
F_{\pi} \sim 0( \sqrt{N} ) \hspace{1cm} L_{2k} \sim 0 (N^{2 - 2k} )
\label{lar45}
\eeq
This means that, for large $N$, we can neglect all even terms except the 
lowest one. Keeping only the leading terms in the large $N$ expansion we 
arrive at the following Lagrangian:
\beq
L = L_0 (U) + \frac{i}{2} q(x) Tr \left[\log U - \log U^{-1} \right]  + 
\frac{1}{a F_{\pi}^{2}} q^2 + \frac{F_{\pi}}{2 \sqrt{2}} 
Tr \left(M U + M^{\dagger} U^{\dagger} \right) - \theta q 
\label{effe454}
\eeq
where $a$ is an arbitrary parameter that is $0(1/N)$ for large $N$ and we
have also allowed for an arbitrary $\theta$ parameter.

It is now convenient to use the algebraic equation of motion for $q(x)$ to
bring eq.(\ref{effe454}) in the following form~\footnote{The resolution of
the $U(1)$ problem in the framework of the large $N$ expansion was given
in Ref.~\cite{WITTEN7,VENEZIA}. See also Ref.~\cite{DIVE}. The effective 
Lagrangian in eq.(\ref{finla43}) was derived in 
Refs.~\cite{ROSE,DIVEVE,WITTEN4}.}:
\beq
L = L_0 (U) + \frac{F_{\pi}}{2 \sqrt{2}} 
Tr \left( M U + M^{\dagger} U^{\dagger} \right) - \frac{a F_{\pi}^{2}}{4}
\left[\theta - \frac{i}{2} Tr \left( \log U - \log U^{-1} \right) \right]^2
\label{finla43}
\eeq
Since $U U^{\dagger}$ is proportional to the unit matrix and the mass matrix
is diagonal we can take the vacuum expectation value of $U$ to be of the following form:
\beq
< U_{ij} > = e^{-i \phi_i} \delta_{ij}  \frac{F_{\pi}}{\sqrt{2}}
\label{vev54}
\eeq
where the angles $\phi_i$ are determined imposing that $<U_{ij} >$ minimizes the
energy corresponding to the Lagrangian in eq.(\ref{finla43}) that is given by:
\beq
E = \frac{aF_{\pi}^2}{4} ( \theta - \sum_i \phi_i )^2 - \frac{F_{\pi}^{2}}{2}
\sum_i \mu_{i}^{2} \cos \phi_i
\label{ene54}
\eeq
after having used eq.(\ref{vev54}). Hence they must satisfy the following
equations:
\beq
\mu_{i}^{2} \sin \phi_i = a ( \theta - \sum_i \phi_i )
\label{min87}
\eeq
It is convenient to work with a field V whose vacuum expectation value is
proportional to the unit matrix. In terms of $U$ it is given by:
\beq
V_{ij} = U_{ik} < U_{kj} >^{-1} \frac{F_{\pi}}{\sqrt{2}}
\label{vij8}
\eeq
and the Lagrangian in eq.(\ref{finla43}) becomes:
\[
L = L_0 (V) + \frac{a F_{\pi}^{2}}{16} \left[Tr ( \log V - \log V^{\dagger}) 
\right]^2 
+ \frac{F_{\pi}}{2 \sqrt{2}} Tr \left[M(\theta) (V + V^{\dagger} - 
\sqrt{2} F_{\pi} ) \right] + 
\]
\beq
+ i \frac{a F_{\pi}}{2 \sqrt{2}} 
( \theta - \sum_i \phi_i ) Tr \left[\frac{F_{\pi}}{\sqrt{2}} 
( \log V - \log V^{\dagger}) - (V - V^{\dagger} ) \right]
\label{newla45}
\eeq
where an inessential constant has been omitted and
\beq
M_{ij} (\theta) = \mu_{i}^{2}(\theta) \delta_{ij} \hspace{2cm} 
\mu_{i}^{2}(\theta) = \mu_{i}^{2} \cos \phi_i
\label{mu89}
\eeq 
Since we are  interested only in those results that follow from current algebra
we can take as we have done in eq.(\ref{u786})
\beq
V = \frac{F_{\pi}}{\sqrt{2}} e^{ i \sqrt{2} \Phi /F_{\pi}} \hspace{2cm}
\Phi = \Pi^{i} \lambda^{i} + \frac{S}{\sqrt{N_f}}
\label{nonli65}
\eeq
In this case one gets the following final Lagrangian:
\[
L = \frac{1}{2} Tr ( \partial_{\mu} V \partial_{\mu} V^{\dagger} ) - 
\frac{1}{2} a N_f S^2 +
\]
\beq
+ \frac{F_{\pi}^{2}}{2} Tr \left[ M(\theta) ( \cos \frac{\sqrt{2}}{F_{\pi}}
\Phi -1) \right] + \frac{ a F_{\pi}}{\sqrt{2}} ( \theta - \sum_i \phi_i)
 Tr \left[ \frac{F_{\pi}}{\sqrt{2}}  \sin \frac{\sqrt{2}}{F_{\pi}}
\Phi -\Phi  \right] 
\label{finla510}
\eeq
where the term implied by the axial anomaly has generated a mass term for the 
singlet field $S$ that has a coefficient $0(1/N)$ for large $N$ 
$(a \sim 0(1/N))$.  

The mass spectrum of the pseudoscalar mesons can just be obtained from the
quadratic part of the Lagrangian:
\beq
L_2 = \frac{1}{2} Tr \left(\partial_{\mu} \Phi \partial_{\mu} \Phi  \right)
- \frac{a}{2} Tr (\Phi )Tr (\Phi ) - \frac{1}{2} Tr \left[ M( \theta) 
\Phi^2 \right]
\label{l243}
\eeq
If we decompose the matrix $\Phi$ as follows:
\beq
\Phi_{ij} = v_i \delta_{ij} + {\tilde{\Pi}}^{\alpha \beta} 
{\tilde{\lambda}}^{\alpha \beta}_{ij}
\label{seo87}
\eeq 
where the matrices ${\tilde{\lambda}}^{\alpha \beta}_{ij}$ are the $N_f ( N_f
-1)$ generators of $SU(N_f )$ that do not belong to the Cartan subalgebra and
we  insert it in eq.(\ref{l243}) we get the 
following two-point functions:
\beq
< {\tilde{\Pi}}^{\alpha \beta} (x) {\tilde{\Pi}}^{\gamma \delta} (y) >^{F.T}=
i \frac{\delta^{\alpha \gamma}{\beta \delta}}{p^2 - M^{2}_{\alpha \beta} 
(\theta) } \hspace{2cm} M^{2}_{\alpha \beta} (\theta) = \frac{1}{2} 
\left(\mu_{\alpha}^{2} (\theta) + \mu_{\beta}^{2} ( \theta) \right)
\label{propa}
\eeq
and
\beq
< v_i (x) v_{j} (y) >^{F.T.} = i A_{ij}^{-1} (p^2)
\label{propa43}
\eeq
where F.T. stands for Fouries transform, the matrix $A_{ij}^{-1}$ is the
inverse of the following matrix:
\beq
A_{ij} (p^2 ) = ( p^2 - \mu_{i}^{2} (\theta) ) \delta_{ij} - a B_{ij}
\label{aij}
\eeq 
and $B$ is a matrix having all elements equal to $1$. The masses of the
physical states can be obtained diagonalizing the mass matrix and are given 
by the following identity:
\beq
det A = \prod_{i=1}^{N_f} ( p^2 - M_{i}^{2} ( \theta) ) = \prod_{i=1}^{N_f}
( p^2 - \mu_{i}^{2} ) \left[1 - a \sum_{j=1}^{N_f} \frac{1}{ p^2 - \mu_{j}^{2}
( \theta)} \right]
\label{detA4}
\eeq
In the chiral limit ($\mu_{i} \rightarrow 0$) one gets $N_{f}^{2} -1$ Goldstone
bosons ($M_{i}=0 $) as expected from the spontaneous breaking of the chiral
symmetry and one particle with mass:
\beq
M_{S}^{2} = a N_f
\label{singlet}
\eeq
Since $ a \sim 1/N$ we see that the mass of the singlet is governed 
in the large-$N$ limit by the 
same factor $N_f /N$ as the coefficient of the  axial anomaly in 
eq.(\ref{axiano}). Therefore we see that the 
resolution of the $U(1)$ problem is intimately related to the existence of
a non vanishing axial anomaly.
\par
A numerical comparison of the spectrum predicted by the mass formula given
in eq.(\ref{detA4}) with the experimental values of the pseudoscalar masses has
been done in Ref.~\cite{VENEZIA} in the case of three flavours. In the limit
where $\mu_1 , \mu_2 << \mu_3 $ one gets the following masses for the $\eta$
and $\eta'$:
\beq
M_{\pm}^{2} = m_{K}^{2} + \frac{3}{2} a \pm \frac{1}{2}  
\sqrt{( 2 M_{K}^{2} - 2 m_{\pi}^{2} -a )^2 + 8 a^2} 
\label{mass32}
\eeq
and the following mixing angle:
\beq
\tan \phi = \sqrt{2} - \frac{3}{2 \sqrt{2}} 
\frac{m_{\eta}^{2} - m_{\pi}^{2}}{m_{K}^{2} - m_{\pi}^{2}}
\label{mix}
\eeq
defined by the relation
\beq
| \eta > = \cos \phi |8> + \sin \phi |1>
\label{mixan}
\eeq
From eq.(\ref{mass32}) we can use the masses of $\eta$ and $\eta'$ to determine
the parameter $a$. We get $a \sim 0.24 (GeV)^2$. Using this value for $a$ and
neglecting the square term in the square root in eq.(\ref{mass32}) one gets:
\beq
m_{\eta}^{2} \simeq m_{K}^{2} + \frac{3 - 2 \sqrt{2}}{2} a = 0.27 (GeV)^2
\label{meta}
\eeq
and
\beq
m_{\eta '}^{2} \simeq m_{K}^{2} + \frac{3 + 2 \sqrt{2}}{2} a = 0.95 (GeV)^2
\label{meta'}
\eeq
that are very close to the experimental values given respectively by 
$0.30 (GeV)^2$ and $0.92 (GeV)^2$. One gets also $\phi = 14^{o}$ that is very
close to the experimental value $\phi = 11^{o}$. The phenomenological 
Lagrangian that we have used give values for the masses that are in good
agreement with the experimental ones. The resolution of the $U(1)$ problem
implies that the parameter $a$ must be different from zero. It can be computed
in pure Yang-Mills theory by computing the following correlator:
\beq
\chi_t \equiv -i \int d^4 y < q(x) q(y) >_{Y.M.} = \frac{1}{2} a F_{\pi}^{2}
\label{topo}
\eeq
that in the literature is known as the topological susceptibility. From the
value of $a$ obtained from the spectrum of pseudoscalar mesons we get the
following value for the topological susceptibility:
\beq
\chi_t  = (180 MeV )^4
\label{top98}
\eeq
Lattice calculation have confirmed this result~\cite{DIGIACOMO}.

At the end of this section we want to discuss the $\theta$ dependence that
follows from the effective Lagrangian in eq.(\ref{finla510}). We start noticing
that, if we consider the Lagrangian in eq.(\ref{effe454}),  we neglect the
terms that come from the fermions and that therefore depend on $U$ and we add a 
source
term $(- i J q)$, we have a Lagrangian that has precisely the same structure
as the corresponding one for the $CP^{N-1}$  model given in eq.(\ref{theta3}).
This means that also in this case we get
\beq
Z( J, \theta) \equiv e^{-i W(J,\theta)} 
= e^{-i a F_{\pi}^{2} (\theta + i J )^2/4} 
\label{zeta38}
\eeq
From it we can compute the vacuum energy:
\beq
E( \theta)  = W(0,\theta) = \frac{a F_{\pi}^{2}}{4} \theta^2
\label{etheta3}
\eeq
and of course the topological susceptibility given in eq.(\ref{topo}).
Using eq.(\ref{singlet}) we can recast eq.(\ref{etheta3}) in the form:
\beq
M_{S}^{2} = \frac{2 N_f}{F_{\pi}^{2}} \frac{d^2 E(\theta)}{d \theta^2}
|_{\theta =0}
\label{witrel}
\eeq
that is the famous Witten's relation~\cite{WITTEN7}. 

The dependence on the $\theta$ parameter for the various physical quantities
is obtained by solving  eqs.(\ref{min87}) that minimize the vacuum energy 
in eq.(\ref{ene54}). They cannot be solved in general analytically. They imply
that  physics is periodic in $\theta$ with period equal to $2 \pi$. In fact,
if $\phi_i = \phi_{i}^{(0)} (\theta, \mu_i ,a )$ is a solution of 
eq.(\ref{min87}), then, for $\theta \rightarrow \theta + 2 \pi$, the solution
can be taken for instance to be of the following form:
\beq
\phi_1 (\theta +2 \pi ) = \phi_i + 2 \pi \hspace{1cm}
\phi_i (\theta + 2 \pi ) = \phi_i \hspace{1cm} i \neq 1
\label{newso98}
\eeq
with no effect on the physics because the physical quantities depend on 
$e^{\pm i \phi_i }$. This means in particular that the vacuum energy must be
a periodic function of $\theta$
\beq
E ( \theta + 2 \pi ) = E ( \theta)
\label{perio86}
\eeq
However this does not necessarily mean a $2 \pi$ periodicity of each solution 
of eq.(\ref{min87}). In general one needs to shift from a solution to another at
some particular value of $\theta$ (typically at $\theta = \pm \pi$) in order to
keep the minimum energy. This can be seen very clearly for instance in 
the case of two flavours with $\mu_1 = \mu_2 \equiv \mu$ for $a >> \mu$ where 
from  eq.(\ref{min87}) one finds $\theta = 2 \phi$. Inserting it in the 
vacuum energy in eq.(\ref{ene54}) one gets
\beq
E ( \theta ) = - F_{\pi}^{2} \mu^2 \sqrt{\frac{1 + \cos \theta}{2}}
\label{ethe4}
\eeq
This shows that at $\theta = \pi$ we shift from a solution to another solution
in order to minimize the energy. Because of this the vacuum energy is periodic
with period $2 \pi$ and not $4 \pi$!!

Let us consider now the case of one flavour and assume instead that the 
quantity $x \equiv a/\mu^2$ is very small. In this limit eq.(\ref{min87}) can
be solved as a power expansion in $x$ and we get~\cite{WITTEN7}:
\beq
\phi = 2 \pi k + x ( \theta - 2 \pi k) + 0( 2 \pi kx)^2
\label{phi85}
\eeq
where $k$ is an integer such that $(2 \pi k x)^2$ is small. But since $a \sim
1/N$ for large $N$ more and more values of $k$ are allowed. For very large $N$
the number of allowed values of $k$ is proportional to $N$. For each value of
$\theta$ only one value of $k$ is the true vacuum, the others correspond to 
metastable states. If we insert the solution given in eq.(\ref{phi85}) in 
eq.(\ref{ene54})  we get for small $x$:
\beq
E (\theta) = \frac{a F_{\pi}^{2}}{4} Min_k ( \theta - 2 \pi k)^2
\label{ethet3}
\eeq
For $- \pi < \theta< \pi$, $k=0$ corresponds to the true vacuum, but for
$\theta > \pi$ then the value $k=1$ corresponds to the true vacuum and so on.
Extracting a $N^2$ factor required by large $N$ counting we can rewrite 
eq.(\ref{ethet3}) as follows:
\beq
E (\theta) = N^2 \frac{a F_{\pi}^{2}}{4} Min_k \left( 
\frac{\theta - 2 \pi k}{N} \right)^2
\label{ethet26}
\eeq
In conclusion we have found that the vacuum energy must be a periodic 
function of $\theta$ (see
eq.(\ref{perio86})) and at the same time must be of the form:
\beq
E(\theta) = N^2 C F((\theta - 2 \pi k)/N ) \hspace{2cm} F(x)= x^2
\label{ethe43}
\eeq
This is only possible if we get a multibranched solution of eq.(\ref{min87}).

\sect{Anti De Sitter space }
\label{AdS}
\vskip 0.5cm

In this section we give some detail about anti De Sitter space in $D \equiv
n+1$ dimensions. De Sitter or anti De Sitter spaces correspond to 
solutions of the pure gravity equations in presence of a cosmological term. 
The action of pure gravity with a cosmological term is given by 
\beq
S = -s \frac{1}{16 \pi G_D} \int d^D x \sqrt{|g|} ( R + \Lambda)
\label{eincos}
\eeq
The factor $s$ in front of the action is $s=1$ if we work with a Minkowski
metric with mostly minus or with a euclidean metric, while $s=-1$ in the case of
a Minkowski metric with mostly plus. From the previous action
we can immediately derive the following eq. of motion:
\beq
R_{\mu \nu} - \frac{1}{2} g_{\mu \nu} R = \frac{1}{2} \Lambda g_{\mu \nu}
\label{eineq}
\eeq
that implies that the scalar curvature is a constant:
\beq
R = \frac{D}{2-D} \Lambda \hspace{3cm} R_{\mu \nu} = \frac{\Lambda}{2 -D} 
g_{\mu  \nu}
\label{concu}
\eeq
De Sitter space corresponds to the case $\Lambda < 0$, while anti De Sitter
space corresponds to the opposite case ($\Lambda >0$).

$AdS_{n+1}$ can be easily represented by embedding it in a flat 
$(n+2)$-dimensional space. Let me call $y^a \equiv ( y^0, y^1, \dots y^n,
y^{n+1} )$ the coordinates of the embedding  space with diagonal metric
equal to $\eta_{ab} = (+1, -1, \dots -1, +1 )$. Then $AdS_{n+1}$ can be shown 
to be the locus characterized by the following equation:
\beq
y^2 = y_{0}^{2} + y_{n+1}^{2} - \sum_{i=1}^{n} y_{i}^{2} = b^2
\label{locus}
\eeq 
where $b$ is a constant anti De Sitter radius. It can be shown that the
previous eq. implies the two eqs. in eq.(\ref{concu}) provided that we
make the following identification:
\beq
\Lambda = \frac{n(n-1)}{b^2}
\label{ide3}
\eeq
Another way of representing $AdS_{n+1}$ is through the stereographic
projection:
\beq
y^0 = \rho \frac{1+x^2}{1 - x^2} \hspace{2cm} y^{\mu} = \rho 
\frac{2 x^{\mu}}{1 - x^2} \hspace{1cm} \mu =1, \dots n+1
\label{ste}
\eeq
where $x^2 = (x^1)^2 +  \dots + (x^n )^2 - (x^{n+1})^2$. According to the
previous transformation we can use the variables $\rho$ and $x^\mu$ instead
of $y^0$ and $y^{\mu}$ to represent the embedding space. Starting from the
flat metric in the embedding space:
\beq
ds^2 = (d y^0)^2 + (d y^{n+1} )^2 - d (\vec{y})^2
\label{flat2}
\eeq
one can rewrite it by using the relations in eq.(\ref{ste}) and one gets
\beq
ds^2 = d\rho^2 - \frac{4 \rho^2}{(1 - x^2)^2} (dx)^2
\label{met3}
\eeq
For fixed $\rho =b$  we get the metric of $AdS_{n+1}$:
\beq
g_{\mu \nu} =  \frac{4 b^2}{(1 - x^2)^2} \eta_{\mu \nu}
\label{met4}
\eeq
where the metric $\eta^{\mu \nu}$ is with mostly plus.

Another parametrization of $AdS_{n+1}$ is the one that appears in 
the near horizon limit of a D $3$-brane. If we work in Minkowski space
we can introduce the variables 
\beq
u = y^0 + i y^{n+1} \hspace{2cm} v = y^0 - i y^{n+1}
\label{ligcon1}
\eeq
while in euclidean space, corresponding to changing the sign
in front of the term $(y^{n+1})^2$, 
we can introduce the alternative variables:
\beq
u = y^0 + y^{n+1} \hspace{2cm} v = y^0 - y^{n+1}
\label{ligcon}
\eeq
In both cases one can rewrite eq.(\ref{locus}) as follows:
\beq
y^2 = uv - \vec{y}^2 = b^2
\label{locus2}
\eeq
Introducing the new variables $ b~\xi^{\alpha}= y^{\alpha}/u$ for $\alpha =1,
\dots, n$ and inserting it in eq.(\ref{locus2}) we can extract $v$ as a
function of $\xi$ and $u$:
\beq
v = b^2 \left( \vec{\xi}^2 u + \frac{1}{u} \right)
\label{vfuu}
\eeq
Then from the flat embedding metric in eq.(\ref{flat2}) written in terms of the
variables $u$ and $v$ we get:
\beq
(ds)^2 = b^2\left( \frac{du^2}{u^2}  + u^2 d \vec{\xi}^2 \right)
\label{mamet}
\eeq
after having used the following equations:
\beq
dv = 2 b^2 \vec{\xi} \cdot d \vec{\xi} u + b^2 \vec{\xi}^2 u du - 
\frac{b^2}{u^2} du
\hspace{2cm} d \vec{y} = bu d\vec{\xi} + b \vec{\xi} du
\label{rel7}
\eeq
Anti De Sitter space has a boundary that is obtained by rescaling the variables:
\beq
y^{\alpha} \rightarrow R {\tilde{y}}^{\alpha} \hspace{1cm} 
u  \rightarrow R {\tilde{u}} \hspace{1cm} 
v^{\alpha} \rightarrow R {\tilde{v}} 
\label{resca}
\eeq
with $R > 0$ and by taking $R \rightarrow \infty$. In this way we get that
the boundary is the manifold satisfying the eq.
\beq
{\tilde{u}}{\tilde{v}} - \vec{{\tilde{y}}}^2 = 0
\label{bou87}
\eeq
But since $tR$ is as good as $R$ the boundary will be described by the two
equations:
\beq
uv - \vec{y}^2 =0 \hspace{2cm} ( u, v, \vec{y} ) \sim ( tu , tv, t \vec{y})
\label{boun76}
\eeq
with $t>0$. We can drop the second condition by just choosing $t$ in such a 
way that
\beq
\vec{y}^2 = 1 = uv = y_{0}^{2} + y_{n+1}^{2}
\label{bou64}
\eeq
This means that the boundary has the topology of $S^1 \times S^{n-1}$, that
is the same topology of compactified Minkowski space with euclidean
compactified time. The usual Minkowski space is recovered when we uncompactify
the two spheres. 

In the last part of this section we introduce additional parametrizations of
$AdS$ space. The first one is obtained  by introducing the
coordinates $(z, \vec{x} ) = (1/u, \vec{\xi} )$. In these coordinates the
AdS metric in eq.(\ref{mamet}) becomes
\beq
ds^2 = b^2 \frac{ dz^2 + d {\vec{x}}^2}{z^2}
\label{met41}
\eeq
The second one corresponds to the cylinder coordinates  defined by the 
following eqs.:
\[
y_{n+1}= b \cosh \rho \cos \tau \hspace{1cm}; \hspace{1cm}
y_0 = b \cosh \rho \sin \tau
\]
\beq
y_i = b \sinh \rho e_{i}
\label{cylcoo}
\eeq
where $e_i$ stands for  a unit vector in $n$ dimensions.
In terms of the previous variables one gets the following metric for 
$AdS_{n+1}$:
\beq
ds^2 = - \cosh^2 \rho d \tau^2 + d\rho^2 + \sinh^2 \rho d\Omega^{2}_{n-1}
\label{cylcoo2}
\eeq
in Minkowski space and
\beq
ds^2 = d\rho^2 + \sinh^2 \rho d\Omega^{2}_{n}
\label{cyleu}
\eeq
in euclidean space where instead the variables used in the previous equation
are defined by
\beq
y_0 = b \cosh \rho ~~~~~~~,~~~~~~y_i = b \sinh \rho e_i~~~~~~i=1, \dots n+1
\label{eucli64}
\eeq
in terms of the embedding space coordinates.
The cavity coordinates $r$ and $\tau$ are instead obtained from the ones
in eq.(\ref{met41}) through the following relations:
\beq
z = \rho \cos \theta \hspace{1cm}; \hspace{1cm} x_i = \rho \sin \theta e_i
\label{cavcoo}
\eeq
where
\beq
\rho = e^{\tau} \hspace{1cm};\hspace{1cm} \cos \theta = \frac{1 - r^2}{1+ r^2}
\label{cavcoo2}
\eeq
In terms of those coordinates the metric of $AdS_{n+1}$ becomes:
\beq
ds^2 = - b^2 \left(\frac{1+ r^2}{1 - r^2} \right)^2 d\tau^2 + 
\frac{4b^2 }{(1-r^2)^2}
\left[dr^2 + r^2 d\Omega_{n-1}^{2} \right]
\label{cav43}
\eeq
An interesting property of $AdS$ space is that a light ray can reach its
boundary in finite time. In fact from the metric in eq.(\ref{cav43}) a light ray
is characterized by the eq.
\beq
\frac{dr}{d \tau} = \frac{1+r^2}{2}
\label{lightray}
\eeq
Integrating the previous equation from the points $(r=0, \tau=0)$ and
$(r=1, T)$ corresponding respectively to the center of $AdS$ space at $\tau=0$
and its boundary at $\tau =T$ we get:
\beq
T = 2 \int_{0}^{1} \frac{dr}{1 +r^2} = \pi/2
\label{time4}
\eeq
This means that a light ray starting from the center of $AdS$ space reaches 
its boundary and comes back to the center in a time interval equal to $\pi$.

\sect{${\cal{N}}=4$ super Yang-Mills and type IIB string }
\label{n=4}
\vskip 0.5cm

In this section we review the main properties of both ${\cal{N}}=4$ super 
Yang-Mills theory and the type IIB string. 

In order to derive the Lagrangian of ${\cal{N}}=4$ super Yang-Mills it is
convenient to start from ${\cal{N}}=1$ super Yang-Mills in ten dimensions and
then dimensionally reduce it to four dimensions. If we do this we obtain the 
following Lagrangian for ${\cal{N}}=4$ super Yang-Mills:
\[
{\cal{L}}=  - \frac{1}{4} F^{a}_{\mu \nu} F^{a \, \mu \nu}
 + \frac{1}{2} \sum_{i=1}^{3} \left( D_{\mu} A_{i} \right)^{a} 
\left( D^{\mu} A_{i} 
\right)^{a} + \frac{1}{2} \sum_{i=1}^{3} \left( D_{\mu} B_{i} \right)^{a} 
\left( D^{\mu} B_{i} \right)^{a} - V(A_i , B_j) + 
\]
\beq
- \frac{i}{2} \left( {\bar{\psi}} \right)^a \gamma^{\mu} \left( D_{\mu} \psi 
\right)^a - \frac{g}{2}  
f^{abc} {\bar{\psi}}^{a} \alpha^{i} A^{b \, i} \psi^{c} - i 
\frac{g}{2} 
f^{abc} {\bar{\psi}}^{a} \beta^{j} \gamma_5 B^{b \, j} \psi^{c} 
\label{neq4sYM}
\eeq
where the potential is equal to:
\beq
V( A_i , B_j ) = \frac{g^2}{4} f^{abc} A_{i}^{b} A_{j}^{c}
f^{afg} A_{i}^{f} A_{j}^{g} + \frac{g^2}{4} f^{abc} B_{i}^{b} B_{j}^{c}
f^{afg} B_{i}^{f} B_{j}^{g} + \frac{g^2}{2} f^{abc} A_{i}^{b} B_{j}^{c}
f^{afg} A_{i}^{f} B_{j}^{g}
\label{potn=4}
\eeq
It contains a gluon field, four Majorana spinors and six real scalars. They all
transform according to the adjoint representation of the gauge group. 
The four-dimensional internal 
matrices $\alpha$ and $\beta$ satisfy the following algebra:
\beq
\{\alpha^i , \alpha^j \} = \{\beta^i , \beta^j \} = -2 \delta^{ij} 
\hspace{1cm} [ \alpha^i , \beta^j ] =0
\label{con2}
\eeq
and
\beq
[ \alpha^i , \alpha^j ] = -2 \epsilon^{ijk} \alpha^{k} \hspace{2cm}
[ \beta^i , \beta^j ] = -2 \epsilon^{ijk} \beta^{k}
\label{excon}
\eeq
They can be chosen to be given by:
\beq
\alpha^i \equiv \eta^{i}_{AB} = \delta_{iA} \delta_{B4} - \delta_{iB} 
\delta_{A4} + \epsilon_{iAB4}
\label{eta}
\eeq
and
\beq
\beta^i \equiv {\bar{\eta}}^{i}_{AB} = \delta_{iA} \delta_{B4} - \delta_{iB} 
\delta_{A4} - \epsilon_{iAB4}
\label{etabar}
\eeq
where $A,B$ are four-dimensional indices and $g \equiv g_{YM}$.

${\cal{N}}=4$ super Yang-Mills is invariant under an internal $SU(4)$ symmetry
group that is an R-symmetry. In order to manifestly see this invariance it is
convenient to introduce the field:
\beq
{\bar{\Phi}}_{AB} = \frac{1}{2 \sqrt{2}} \left[\eta^{i}_{AB} A_{i} - 
{\bar{\eta}}^{i}_{AB} B_{i} \right]
\label{Phi2}
\eeq
satisfying the condition:
\beq
\Phi^{AB} \equiv \frac{1}{2} \epsilon^{ABCD} {\bar{\Phi}}_{CD} = \Phi_{AB}^{*}
\label{phi2}
\eeq
This antisymmetric field transforms according to the vector $6$ representation
of $SO(6)$ that has the same algebra as $SU(4)$. By rewriting the Lagrangian in 
eq.(\ref{neq4sYM}) in terms of $\Phi$ and in terms of the Weyl spinors, that 
transform
according to the $4$ representation of $SU(4)$, one gets~\footnote{This form of
the Lagrangian has been written together with F. Fucito and G. Travaglini.}:
\[
L= - \frac{1}{4} F_{\mu \nu}^{a} F^{\mu \nu}_{a} + 
(D_{\mu} {\bar{\Phi}}_{AB})_a (D^{\mu} \Phi^{AB})_a  - i \psi_{a}^{\alpha A}
\sigma_{\alpha \dot{\alpha}}^{\mu} (D_{\mu} {\bar{\psi}}^{\dot{\alpha}}_{A})_a
+
\]
\beq
- g^2 f^{abc}\Phi_{b}^{AB} \Phi_{c}^{CD} f^{ade} {\bar{\Phi}}^{d}_{AB} 
{\bar{\Phi}}^{e}_{CD} - g \sqrt{2} f^{abc} \left[\psi_{a}^{\alpha A} 
{\bar{\Phi}}_{AB}^{b}
\psi_{\alpha}^{c B} + {\bar{\psi}}^{a}_{\dot{\alpha} A} \Phi^{AB}_{b} 
{\bar{\psi}}^{\dot{\alpha}}_{cB} \right]
\label{maninv}
\eeq
It is manifestly invariant under the $SU(4)$ R-symmetry
transformations:
\beq
\psi_{\alpha}^{A} \rightarrow U^{A}_{\,\,\,B} \psi_{\alpha}^{B}
\hspace{1cm}
{\bar{\psi}}_{{\dot{\alpha}} A} \rightarrow ({U^{*}})_{A}^{\,\,\,B} 
{\bar{\psi}}_{{\dot{\alpha}}B}
\label{tra31}
\eeq
and
\beq
\Phi^{AB} \rightarrow U^{A}_{\,\,\,C} \Phi^{CD} ( U^{T})_{D}^{\,\,\,B}
\hspace{1cm}
{\bar{\Phi}}_{AB} \rightarrow (U^{*} )_{A}^{\,\,\,C} {\bar{\Phi}}_{CD} 
( U^{\dagger})^{D}_{\,\,\,B}
\label{tra32}
\eeq
where $U$ is a unitary matrix $( U U^{\dagger} =1 )$.

Lagrangian in eq.(\ref{neq4sYM}) can also be written in the ${\cal{N}}=1$ 
superfield formalism. One must introduce three chiral superfields $\Phi_i$ 
 and one obtains the following Lagrangian:
\[
{\cal{L}} = \int d^2 \theta d^2 {\bar{\theta}}
\sum_{i=1}^{3} {\bar{\Phi}}_{i} {\rm e}^{2g V} \Phi_{i} + \frac{1}{8 \pi} Im 
\left[ \int d^2 \theta \, \tau W^{\alpha} W_{\alpha} \right]
\]
\beq
+ \left[ \int d^2 \theta  \sqrt{2} g f^{abc} \Phi_{1}^{a} \Phi_{2}^{b} 
\Phi_{3}^{c}  + h.c. \right]
\label{lan=4}
\eeq
${\cal{N}}=4$ super Yang-Mills has  no trace 
anomaly in the sense that the trace of the energy-momentum tensor is zero
also in the full quantum theory. It is a finite quantum field theory.
As we have seen above it is also invariant under a $SU(4)$ R-invariance. This 
symmetry is not manifest
in the formulation with ${\cal{N}}=1$ superfields. Only a $SU(3) \times U(1)$ 
symmetry is  manifest in this formulation. The $SU(3)$ corresponds to a $SU(3)$
rotation of the three superfields $\Phi_i$, while the $U(1)$ acts on the four
superfields as:
\beq
\Phi_i (\theta) \rightarrow e^{i 2\alpha/3} \Phi_{i} ( \theta e^{-i \alpha} )
\hspace{1cm} W_{\alpha}( \theta) \rightarrow e^{i \alpha} W_{\alpha} (\theta 
e^{- i \alpha}) 
\label{Rsy} 
\eeq
According to the previous eqs. the fermions of the three superfields $\Phi_i$
have chiral weight equal to $- 1/3$, while the fermion of the superfield 
$W_{\alpha}$ has chiral weight equal to $1$. This means that the sum of their
chiral weights is vanishing implying that ${\cal{N}}=4$ super Yang-Mills has
no $U(1)$ axial anomaly.
Finally the action in eq.(\ref{neq4sYM}) is also invariant under 
conformal supersymmetry transformations. They  can be
obtained by dimensionally reducing the ${\cal{N}}=1$ supersymmetry
transformations in ten dimensions. Since the spinor in ten dimensions is
a Weyl-Majorana spinor the ten-dimensional theory is invariant under $16$
supersymmetries. On the other hand the
four-dimensional quantum theory is  conformal invariant and therefore it is 
also invariant under the same supersymmetry transformations as before, but with
a supersymmetry parameter $\alpha = \gamma_{\mu} x^{\mu} \beta$ that is
space-time dependent, while $\beta$ is space-time independent. They correspond
to additional $16$ supersymmetries and therefore we conclude that ${\cal{N}}=4$
super Yang-Mills in four dimensions is invariant under $32$ supersymmetries.

Finally there is a very strong evidence  that ${\cal{N}}=4$ super 
Yang-Mills is invariant under $SL(2,Z)$ transformations that act on the complex 
coupling constant $\tau$ defined in terms of the gauge coupling constant and
the $\theta$-parameter as:
\beq
\tau \rightarrow \tau ' = \frac{a \tau + b}{c \tau + d} \hspace{2cm}
\tau = \frac{\theta}{2 \pi } + i \frac{4 \pi}{g_{YM}^{2}}
\label{mod}
\eeq
where $a, b, c$ and $d$ are integers satisfying the condition $ad -bc =1$.
For $\theta =0$ the transformation in eq.(\ref{mod}) relates weak with strong 
coupling. The invariance under $SL(2,Z)$  is a precise way of stating that 
this theory satisfies the Montonen-Olive~\cite{MONTOLI}
duality in the sense that it can be equivalently formulated as a theory of
fundamental $W$-mesons having magnetic monopoles  as solitons or as a theory
of fundamental magnetic monopoles with the $W$-mesons appearing as solitons,
the two formulations having essentially the same Lagrangian.

Type IIB string is  a theory of closed superstrings
involving both right and left movers. The right and left spinors in the R sector
have the same chirality. Therefore it is a chiral  theory with no gauge and
gravitational anomalies. It is a ${\cal{N}}=2$ supersymmetric theory in ten
dimensions; this means that it contains $32$ supersymmetry charges. These
supersymmetries are kept also when one compactifies it  on the
background $AdS_5 \times S_5$. In the massless bosonic sector the ten 
dimensional 
theory contains a graviton $g_{\mu \nu}$, a dilaton $\phi$, an axion field
$\chi$, two 2-form potentials $B_{\mu \nu}^{(1)}$ and $B_{\mu \nu}^{(2)}$
and a 4-form potential $A_{\mu \nu \rho \sigma}$ with self-dual field
strenght, while the massless fermionic sector consists of two gravitinos
and two dilatinos having both the same chirality. If we forget for a moment
the R-R self-dual $5$-form field the low-energy effective
Lagrangian for type IIB theory has the following form:
\beq
S_{IIB} = \frac{1}{ 2 \kappa^2} \int d^{10} x \sqrt{-g} \left[ R + 
\frac{1}{4} Tr( \partial {\cal{M}} \partial {\cal{M}}^{-1}) - 
\frac{1}{12} H^{T} {\cal{M}} H \right]
\label{IIB}
\eeq
where we have combined the field strenghts $H^{(1)}$, corresponding to the NS-NS
potential, and $H^{(2)}$, corresponding to the R-R potential, in a 
two-component vector
$H = dB$ and  the two scalar fields in the symmetric $SL(2,R)$
matrix:
\beq
{\cal{M}} = e^{\phi} \left( \begin{array}{cc} |\lambda|^2 & \chi \\
                                           \chi & 1 
                          \end{array} \right)
\label{sl2rma}
\eeq
with
\beq
\lambda = \chi + i e^{-\phi}
\label{lam}
\eeq
This action is manifestly invariant under the global $SL(2,R)$ transformation:
\beq
{\cal{M}} \rightarrow \Lambda {\cal{M}} \Lambda^{T}
\hspace{2cm}
H \rightarrow (\Lambda^T )^{-1} H \hspace{1cm} \Lambda = \left( 
             \begin{array}{cc} a & b \\
                             c & d \end{array} \right) 
\label{sl2rtra}
\eeq
The metric in the Einstein frame and the 4-form potential are left
invariant by the $SL(2,R)$ transformation. 
In terms of the matrix $\Lambda$ previously defined we get that
\beq
\lambda \rightarrow \frac{a \lambda +b}{c \lambda +d} \hspace{1cm}
H^{(1)} \rightarrow d H^{(1)} - c H^{(2)} \hspace{1cm}
H^{(2)} \rightarrow -b H^{(1)} + a H^{(2)}
\label{tras}
\eeq
In particular the transformation on the matrix ${\cal{M}}$ given in 
eq.(\ref{sl2rtra}) implies that the quantity $\lambda$ defined in 
eq.(\ref{lam}) transforms exactly as $\tau$ in eq.(\ref{mod}).
Although the  low-energy Lagrangian is invariant under a $SL(2,R)$ symmetry 
it can be seen that it reduces to an $SL(2,Z)$ symmetry in the quantum theory 
in order  to have the Dirac quantization condition satisfied. 

In conclusion we have seen that both ${\cal{N}}=4$ super Yang-Mills
and type IIB string theory compactified on $AdS_5 \otimes S_5$ 
have the same symmetries. 

\sect{The Maldacena conjecture}
\label{malda}

In the introduction we have seen how a system of $N$ D $3$-branes is, on the
one hand, a classical
solution of the supergravity equations of motion containing the metric, the
dilaton and the self-dual $5$-form field strenght and, on the other hand, is
described by the Born-Infeld action that at low energy reduces
to the Yang-Mills action in eq.(\ref{biexp}) with gauge group $U(N)$ or more 
precisely to its 
${\cal{N}}=4$ supersymmetric extension. But is it possible to find a more
precise connection or even an equivalence between the world volume 
${\cal{N}} =4$ supersymmetric Yang-Mills theory and supergravity or better
superstring that is a consistent quantum theory? The key to answer this
question came from comparing~\cite{KLEBA1,GUB} the low energy absorption 
cross sections of 
massless bulk fields as graviton and dilaton computed either by using the 
supergravity classical solution or the Born-Infeld action. To great 
surprise it was found that the two calculations exactly 
agree~\cite{KLEBA1,GUB}. Moreover,
if one studies the range of validity of the two previous calculations, it
clearly appears that the one based on supergravity  is expected to give an 
exact information on ${\cal{N}}=4$ super Yang-Mills theory for large 't Hooft
coupling ($\lambda \rightarrow \infty$)~\cite{KLEBA1}.  

These properties of a system of $N$ D $3$-branes together with the 
observation~\cite{MALDA} that it is the region around the throat of metric of
the D $3$-branes that is the fundamental one to connect the supergravity
solution with ${\cal{N}}=4$ super Yang-Mills, brought Maldacena~\cite{MALDA} 
to conjecture that ${\cal{N}}=4$ super Yang-Mills should be somehow equivalent
to type IIB string theory compactified on $AdS_5 \times S^5$ that is in fact
the metric in the throat region. In the following we will briefly sketch
the main lines of his argumentation.

Let us consider the low-energy limit of the Born-Infeld action for a system of
$N$ D $3$-branes and of the bulk supergravity action. Since it
amounts to take the limit $\alpha' \rightarrow 0$ with both $g_s$ and $N$ 
fixed, in this limit we obtain the action of ${\cal{N}}=4$ super Yang-Mills, 
that we have discussed in sect.~\ref{n=4} and that is the low-energy limit of 
the Born-Infeld action together with free gravitons. This is
a consequence of the fact that both the interaction between bulk fields as for
instance the graviton and that between bulk fields and those living on the 
brane as the Yang-Mills fields, being proportional to the Newton's constant 
$\kappa \sim (\alpha')^2$, go to zero when $\alpha ' \rightarrow 0$.

On the other hand if we look at the classical solution in eqs.(\ref{clasol}),
(\ref{clasol2}) and (\ref{acca}) we see that it interpolates between  flat 
ten-dimensional Minkowski metric obtained for $r \rightarrow \infty$ and a 
metric with a long throat obtained in the limit $r \rightarrow 0$. In 
particular for $p=3$ the metric is non singular when $r \rightarrow 0$ and
in this limit becomes  that of $AdS_5 \times S^5$.
More precisely this can be seen by taking  the near-horizon limit of a system
of $N$ D $3$-branes defined by
\beq
r \rightarrow 0 \hspace{2cm} \alpha ' \rightarrow 0 \hspace{2cm} U \equiv
\frac{r}{\alpha '} = fixed
\label{lim}
\eeq
where the Regge slope is taken to zero, while $U$ is kept fixed. In 
this limit we can neglect the factor $1$ in the function $H$ in 
eq.(\ref{acca}) and the metric in eq.(\ref{clasol}) becomes:
\beq
\frac{(ds)^2}{\alpha'} \rightarrow \frac{U^2}{\sqrt{4 \pi N g_{s}}} (dx_{3+1})^2
+ \frac{\sqrt{4 \pi N g_{s}}}{U^2} dU^2 + \sqrt{4 \pi N g_{s}} d \Omega_{5}^{2}
\label{nmet}
\eeq
This is the metric of the manifold $AdS_5 \times S^5 $ where the two radii of
$AdS_5$ and $S^5$ are equal and given by:
\beq
R_{AdS_5}^{2} = R_{S_5}^{2} \equiv b^2 = \alpha' \sqrt{4 \pi N g_{s}} 
\label{rad}
\eeq
that, using the relation $g_{YM}^{2} = 4 \pi g_s$ following from 
eq.(\ref{biexp}) for $p=3$, implies:
\beq
\frac{b^2}{\alpha '} = \sqrt{N g_{YM}^{2}}
\label{rad2}
\eeq
If we have sufficiently soft gravitons (i.e. gravitons with wave lenght much
bigger than the radius of the throat $b$) outside the throat they cannot 
interact with the excitations far down in the throat as it is confirmed by
the fact that their absorption cross-section  is vanishing at low energy.  
On the other hand a string excitation far down inside the
the throat, although its proper energy (the energy measured in the reference 
frame instantaneously at rest at $r$) diverges at low energy ($\alpha ' 
\rightarrow 0)$, being proportional to $ E_p 
\sim 1/\sqrt{\alpha '}$, is not negligible because its energy measured in 
the frame of reference where the time is the one appearing in the first term
of the r.h.s. of eq.(\ref{nmet})  is given by:
\beq
E_t \sim \frac{r}{b} E_p \sim \frac{r}{b \sqrt{\alpha'}} \sim \frac{r}{\alpha'}
= U
\label{et}
\eeq 
that is kept fixed in the limit $\alpha ' \rightarrow 0$. Therefore from the
point of view of the classical solution we are left with free gravitons and
all the string excitations living far down inside the throat that are 
described by
type IIB string theory compactified on $AdS_5 \times S^5$. By comparing this
result with the one obtained from the Born-Infeld action Maldacena 
 has formulated the conjecture that ${\cal{N}} =4$ super Yang-Mills is
equivalent to type IIB string theory compactified on $AdS_5 \times S^5$.
The precise relation between the parameters of the gauge and string theory
is given in eq.(\ref{rad}), where $N$ is equal to the number of colours in the
gauge theory and to the flux of the $5$-form field strenght in the supergravity
solution. Since the classical solution in eq.(\ref{nmet}) is a good 
approximation when the radii of $AdS_5$ and $S^5$ are very big 
\beq
\frac{b^2}{\alpha'} >> 1 \Longrightarrow Ng_{YM}^{2} \equiv \lambda >>1~,
\label{bigrad}
\eeq
in the strong coupling limit of the gauge theory we can restrict ourselves
to the type IIB supergravity compactified on $AdS_5 \otimes S^5$.


In conclusion, according to the Maldacena conjecture, classical supergravity
is a good approximation if $\lambda >>1$, while in the 't Hooft limit in which
$\lambda$ is kept fixed for $N \rightarrow \infty$ classical string theory
is a good approximation for  ${\cal{N}}=4$ super Yang-Mills. In the 't Hooft
limit in fact string loop corrections are negligible ($g_s << 1$) 
as it follows from the equation: $ \lambda = 4 \pi g_s N$ for $\lambda$ fixed
and $N \rightarrow \infty$. Finally Yang-Mills
perturbation theory is a good approximation when $\lambda <<1$.

The strongest evidence for the validity of the Maldacena conjecture comes from
the fact that both ${\cal{N}}=4$ super Yang-Mills and type IIB string 
compactified on $AdS_5 \otimes S^5$ have the same symmetries. They are, in 
fact, both invariant under $32$ supersymmetries, under the conformal group
$O(4,2)$, corresponding to the isometries of $AdS_5$, under the $R$-symmetry
group $SU(4)$, corresponding to the isometries of $S^5$ and under the 
Montonen-Olive duality~\cite{MONTOLI} based on the group $SL(2,Z)$. 

If the Maldacena conjecture is true, as as it seems to be implied by the 
many positive checks of its validity, then this is the first time that a 
string theory is recognized to come out from a gauge theory. In particular
it is important to stress that this does not contradict the fact mentioned 
earlier that a string theory contains gravity while the gauge theory does not,
because in this case the two theories live in different spaces: IIB string 
theory lives on 
$AdS_5 \otimes S_5$, while ${\cal{N}} =4$ super Yang-Mills lives on the
boundary of $AdS_5$ that is our four-dimensional Minkowski space. The 
equivalence between the two previous theories realizes the holographic 
idea~\cite{HOOF,SUSS}
that a quantum theory of gravity is supposed to satisfy. A new puzzle,
however, arises in this case because we usually connect a string theory
with a confining gauge theory, while instead ${\cal{N}}=4$ super Yang-Mills
is a conformal invariant theory and therefore is in the Coulomb and not in
the confining phase. The fact that ${\cal{N}}=4$ super Yang-Mills is in a
Coulomb phase is confirmed by the calculation of the Wilson loop where a
Coulomb potential between two test charges is found~\cite{MALDA2}.
\par
If two theories, as the type IIB string theory compactified on 
$AdS_5 \otimes S^5$ and ${\cal{N}}=4 $ super Yang-Mills theory, are 
equivalent then it must be possible to specify for each field $Q(x)$
of the boundary Minkowski theory the corresponding field $\Phi(y)$ of the bulk
string theory 
and to show that, when we compute corresponding correlators in the two theories,
we get the same result. In particular, in the boundary theory one can
easily compute the generating functional for correlators involving  $Q(x)$
\beq
Z(\Phi_0 ) = < e^{\int d^4 x \Phi_0 (x) Q(x)} >
\label{gene1}
\eeq
By taking derivatives with respect to the arbitrary source
$\Phi_0 (x)$ one can compute any correlator involving the boundary field
$Q(x)$. In Refs.~\cite{KLEBA,WITTEN2} the recipy for computing $Z(\Phi_0)$
in the bulk theory has been given. First of all 
one must identify $\Phi_0 (x)$ with the boundary value of the field
$\Phi (y)$, which lives in the bulk theory and that corresponds to the composite
$Q(x)$ of the boundary theory. Then the generating functional given in
eq.(\ref{gene1}) can just be obtained by performing in the bulk theory
the functional integral
over $\Phi$ with the restriction that its boundary value be $\Phi_0$:
\beq
Z( \Phi_0 ) = \int_{\Phi \rightarrow \Phi_0} D \Phi \,\,e^{-S [ \Phi]}
\label{gene2}
\eeq
In computing the previous functional integral we can use classical 
supergravity in the regime where $\lambda >> 1$. Otherwise for an
arbitrary value of fixed $\lambda$ for $N \rightarrow \infty$ we need to
compute
the tree diagrams of type IIB string theory compactified on $AdS_5 \otimes
S^5$. 

A number of bulk fields has been identified to correspond to the various
gauge invariant composite fields of ${\cal{N}}=4$ super Yang-Mills. We
do not discuss  this correspondence in detail here. In the
following we will just describe in some detail the correspondence between
the dilaton field of type IIB supergravity and the composite given by the
Yang-Mills Lagrangian $F^2 \equiv F_{\mu \nu}^{a} F^{a \mu \nu}$ showing
that the two-point functions that one obtains from both 
eqs.(\ref{gene1}) and (\ref{gene2})  are  coincident~\cite{KLEBA,WITTEN2}.

Since ${\cal{N}}=4$ super Yang-Mills theory with gauge group $SU(N)$ is a
conformal invariant quantum theory and the composite field $F^2$ has
dimension $4$ the two-point function involving two $F^2$ fields must have 
the following form:
\beq
< F^2 (x) F^2 (z) >  \sim \frac{N^2}{(\vec{x} -\vec{z})^8}
\label{f2f2}
\eeq    
apart from an overall constant that we do not care to compute. The previous
correlator can also be obtained by using the lowest order perturbation
theory in ${\cal{N}}=4$ super Yang-Mills. $\vec{x}$ denotes here a Minkowski
four-vector.

In the bulk theory we only need the dilaton kinetic term in
type IIB supergravity in $D=10$ compactified on $AdS_5 \otimes S^5$. Taking
into account that the volume of $S^5$ is equal to $\pi^3 b^5$, where $b$ is
given in eq.(\ref{rad}), we need to consider the following action:
\beq
S =\frac{\pi^3 b^5}{4 \kappa^{2}_{10}} \int d^5 x \sqrt{g} g^{\mu \nu}
\partial_{\mu} \Phi \partial_{\nu} \Phi 
\label{act}
\eeq
where $g_{\mu \nu} = \frac{b^2}{z^{2}} \delta_{\mu \nu}$ is the metric
of $AdS_5$ in the so-called Poincar{\'{e}} coordinates given in 
eq.(\ref{met41}). 
In the limit $\lambda >> 1$, where classical supergravity is a good
approximation, we just need to solve the dilaton eq. of motion given by:
\beq
\partial_{\mu} \left[ \sqrt{g} g^{\mu \nu} \partial_{\nu} \Phi \right] =0
\label{eqmo}
\eeq
The solution of the previous equation, that is equal to $\Phi_0$ on the
boundary (corresponding to the limit $z \rightarrow 0$), can be given in 
terms of the Green's function:
\beq
\Phi (z, \vec{x}) = \int d^4 {\vec{x}}\,\, K( z , \vec{x}; \vec{y}  )
\,\,\Phi_0 ( \vec{z}  )
\label{solo2} 
\hspace{.5cm};\hspace{.5cm}
K( z , \vec{x}; \vec{y}  ) \sim \frac{z^{4}}{[z^{2} + ( \vec{x}
- \vec{y}  )^2 ]^4}
\eeq
Inserting the solution found in eq.(\ref{solo2}) in the classical action we get
that the contribution to the classical action is entirely due to the
boundary term
\beq
S = \frac{\pi^3 b^8}{4 \kappa_{10}^{2}} \int d^4 \vec{x} z^{-3} \Phi
\partial_0 \Phi |_{\epsilon}^{\infty} \sim - \frac{\pi^3 b^8}{4 
\kappa_{10}^{2}} \int d^4 \vec{x}\int d^4 \vec{y}
\frac{\Phi_0 (\vec{x}) \Phi_0 (\vec{y} ) }{(\vec{x} - \vec{y}  )^8 }
\label{claact}
\eeq 
where we have introduced a cut off $\epsilon$ at the lower limit of integration,
that, however, cancels out after having inserted  eq.(\ref{solo2}) in
eq.(\ref{claact}).
In conclusion in the classical approximation ($\lambda >>1$) we get
\beq
Z ( \Phi_0 ) = \exp \left[\frac{\pi^3 b^8}{4 \kappa_{10}^{2}} \int d^4 \vec{x}
\int d^4 \vec{x}'
\frac{\Phi_0 (\vec{x}) \Phi_0 (\vec{x}' ) }{(\vec{x} - \vec{x} ' )^8 } \right]
\label{clagen}
\eeq
Taking into account eq.(\ref{rad}) and that $2 \kappa_{10}^{2} = ( 2 \pi)^7 
g_{s}^{2} (\alpha ')^4 $, from the previous equation we can get
immediately the two-point function:
\beq
< F^2 (x) F^2 (z) > =
\frac{\partial^2 Z (\Phi_0 )}{\partial \Phi_0 (\vec{x} ) \partial \Phi_0 
(\vec{z} ) }  \sim \frac{N^2}{(\vec{x} - \vec{z} )^8} 
\label{corre}
\eeq 
that agrees with the expression given in eq.(\ref{f2f2}).
Notice, however, that the supergravity approximation is in general only valid
for large values of $\lambda$ (see eq.(\ref{bigrad})), while  the previous
example shows that it seems to be  valid for any value of $\lambda$. This is, 
of course,
a consequence of the conformal invariance of ${\cal{N}}=4$ super Yang-Mills
that requires the vanishing of the contribution to the two-point function
of all string corrections to the supergravity action. The same result is
also true if one computes the three point function involving $F^2$
or the two and three-point functions involving the energy-momentum tensor.
Actually, using the value of $\kappa_{10}$ given just after eq.(\ref{clagen})
together with eq.(\ref{rad}), it is easy to see that the factor in front of
the Einstein action or of the dilaton kinetic term is proportional to $N^2$
and does not depend on the gauge coupling constant $g_{YM}$. This means that
the pure supergravity approximation will never give a dependence of the
correlators of gauge theory on the gauge coupling constant. In order to
obtain the dependence on the gauge coupling constant we need to add string
corrections. Let us restrict ourselves to the pure gravity part of type IIB
supergravity. The action with the first string correction containing only the 
metric is given by: 
\beq
S = \frac{1}{2 \kappa_{10}^2} \int d^{10} x \sqrt{-g} \left[ 
e^{-2 \phi} R +
\frac{(\alpha')^3}{3 \cdot 2^7} R^4 \cdot e^{-1/2 \phi} f(\tau, {\bar{\tau}})
\right]
\label{strcorr}
\eeq 
where 
\beq
 f (\tau, {\bar{\tau}}) = {\sum_{(n,m)}}' 
\frac{\tau_{2}^{3/2}}{|m+ n \tau|^3}~~~~,~~~~\tau= \chi +i e^{-\phi}~. 
\label{taumo}
\eeq
The prime indicates that the term $(n,m) = (0,0)$ is excluded from the sum,
$\tau_2 \equiv Im \tau$  and
$\phi$ and $\chi$ are respectively the dilaton and the R-R scalar of type IIB
string. The function~\cite{GREEN} $f$ that is invariant under the $SL(2,Z)$ 
transformation $\tau \rightarrow (a \tau +b)/(c \tau + d) $, 
can be expanded for small values of $e^{\phi} = g_s$ getting
\[
e^{-\phi/2} f ( \tau, {\bar{\tau}} ) = 2 \zeta (3) e^{-2 \phi} + 
\frac{2 \pi^2}{3} + 
\]
\beq
+ (4 \pi)^{3/2} e^{- \phi/2} \sum_{M >0} Z_{M}
M^{1/2} \left[ e^{2 \pi i M \tau} + e^{2 \pi i M {\bar{\tau}}}  \right]
(1 + \e^{\phi}/M )
\label{expamo}
\eeq
The first term in the r.h.s. of the previous equation comes from the tree 
string diagrams, the second from one-loop
string corrections, while the rest is the contribution of D instantons.
If one now uses the action in eq.(\ref{strcorr}) to
compute the two and three-point function for the energy-momentum tensor
of ${\cal{N}}=4$ super Yang-Mills one finds that the extra term gives no
contribution, because it identically vanishes~\cite{BANKS} when we compute it 
in the background $AdS_5 \otimes S^5$. But instead it gives a non trivial 
contribution to the four-point amplitude. 
In particular, since the four-point amplitude
will include the function $f$, and since in the $AdS/CFT$ correspondence
the Yang-Mills coupling constant is related to the string coupling constant
through the relation: $g_{YM}^{2} = 4 \pi g_s $ we immediately see from 
eq.(\ref{expamo}) that the D instanton contribution becomes the usual
instanton contribution in Yang-Mills theory~\cite{BANKS}. This result is 
confirmed
by explicit calculations in ${\cal{N}} =4$ super Yang-Mills~\cite{BIANCHI}.  
A quick way of fixing the dependence on $N$ and on the gauge coupling
constant of the two terms present in the action in eq.(\ref{strcorr}) is 
by remembering that under a rescaling of the metric by $b^2$ we get the
following relations:
\beq
g_{\mu \nu} \rightarrow b^2 g_{\mu \nu} \hspace{1cm};\hspace{1cm}
\sqrt{g} \rightarrow b^{10} \sqrt{g} 
\label{resca35}
\eeq
and 
\beq
R \rightarrow b^{-2} R \hspace{1cm};\hspace{1cm}
R^4  \rightarrow b^{-8} R^4
\label{resca33}
\eeq 
By the previous rescalings (where we have also taken into account that the 
integration over the sphere $S^5$ gives an extra factor $b^5$) we get that 
the coefficient of
the Einstein term in eq.(\ref{strcorr}) is proportional to $N^2$ as previously
found, while the coefficient of the first string correction in 
eq.(\ref{strcorr}) is proportional to $N^2 ( N g_{YM}^{2} )^{-3/2}$ that depends
explicitly on the gauge coupling constant.

We have seen that in the equivalence between ${\cal{N}}=4$ super Yang-Mills
and type IIB string theory compactified on $AdS_5 \times S^5$ the Yang-Mills
action $F^2$ that has conformal dimension equal to $4$ corresponds to the
massless dilaton. In Ref.~\cite{WITTEN2} it has been shown that a massive 
scalar field with mass equal to $m$ corresponds in the conformal field theory 
to a composite operator with dimension $\Delta$ equal to
\beq
\Delta = 2 + \sqrt{ 4 + b^2 m^2}
\label{delta2}
\eeq 
Here we will not derive this  result whose derivation can be found in 
Ref.~\cite{WITTEN2}, but we will only derive eq.(\ref{delta2}) in the limit
of very large $b\,\,m$. Let us consider a scalar field $\Phi$ of the bulk theory 
with mass $m$ that corresponds to a composite field $F(x)$ of the boundary
theory with conformal dimension $\Delta$. The two-point function in the 
boundary theory is fixed apart from an overall normalization by the conformal 
invariance of the theory. This means that:
\beq
< F(\vec{x} ) F( \vec{y} ) > \sim [ |\vec{x}  - \vec{y} | \mu ]^{- 2 \Delta}
\label{ff5}
\eeq
where we have made the composite $F(x)$ dimensionless by multiplying it
with a factor $\mu^{- \Delta}$ ($\mu$ is a parameter with  dimension
of a mass). On the other hand the previous two-point function of the boundary
theory is also equal to the two-point function of the bulk theory involving
the corresponding field $\Phi$:
\beq
< F( \vec{x} ) F( \vec{y}) >~~ \sim~~ < \Phi( \vec{x} ) \Phi(\vec{y}) >
\label{ff6}
\eeq 
if the two points $\vec{x}$ and $\vec{y}$ are on the boundary of $AdS$ space,
i.e. in Minkowski four-dimensional space. 
But the propagator of a 
free particle with mass $m$ in the bulk theory for $m$ very large is given
by the particle action computed along a geodesic that connects the two points
$\vec{x}$ and $\vec{y}$. 
The action describing a particle moving in AdS space is given by:
\beq
S = m \int \sqrt{g_{\mu \nu} dx^{\mu} dx^{\nu}} 
\label{act87}
\eeq  
Choosing for the AdS metric the one given in eq.(\ref{met41}) and the two points
of the boundary to be at $x= \pm a$ we get:
\beq
S = mb \int_{-a}^{a} \frac{dx}{z} \sqrt{1 + \left(\frac{dz}{dx} \right)^2}
\label{azin65}
\eeq
The geodesic satisfies the equation:
\beq
\left(\frac{dz}{dx} \right)^2 = \left(\frac{z_0}{z} \right)^2 -1
\label{geo76}
\eeq
where $z_0$ is the minimum value taken by $z$. It is easy to see that the
previous eq. defines a circle with center on the boundary at $x=0$ and with
radius $z_0 = a$, that connects the two points on the boundary at $x= \pm a$.
When we insert the geodesic solution in the original action we get:
\beq
S = 2 mb \int_{\epsilon}^{z_0} \frac{dz}{z} \frac{1}{\sqrt{1- \left(
\frac{z}{z_0} \right)^2}}  
\label{azi54}
\eeq
where we have introduced an infrared cutoff $\epsilon$ in the bulk theory.
By performing the integral in the limit of small $\epsilon$ we get:
\beq
< \Phi (-a) \Phi (a) > \sim e^{- S} \sim e^{- 2 mb \log a/\epsilon} =
\left( \frac{a}{\epsilon} \right)^{- 2 mb}
\label{phiphi5}
\eeq 
From eqs.(\ref{ff5}), (\ref{ff6})  and (\ref{phiphi5}) we get for 
large values of the mass $m$:
\beq
\Delta \sim m b
\label{delt54}
\eeq
that agrees with eq.(\ref{delta2}) for $b\,m >>1$. In addition by comparing
eqs.(\ref{ff5}) and (\ref{phiphi5}) we can see that the infrared cutoff 
$\epsilon$ of the bulk theory corresponds
to an ultraviolet cutoff $\mu = 1/\epsilon$ of the boundary 
theory~\cite{SUSSWI}.

\sect{Finite temperature ${\cal{N}} =4$ super Yang-Mills }
\label{fintemp}

In the previous section we have briefly seen how the Maldacena conjecture 
provides for the first time a very strong evidence for the appearence of a 
string theory in a non-perturbative gauge theory precisely realizing the 
ideas reviewed in sect.~\ref{largeN} on the large $N$ expansion in QCD and 
without running into the problem that a string theory contains gravity while
the gauge theory does not. On the other hand we are immediately confronted with 
a new puzzle because ${\cal{N}}=4$ super Yang-Mills is in the
Coulomb phase and therefore the emergence of a string has nothing to do with
the confining properties of the theory. In order to get a confining theory we
have to get rid of the conformal invariance of the theory. The simplest way
for doing so is by considering ${\cal{N}}=4$ super Yang-Mills at finite
temperature. But, since  bosons have periodic and fermions anti-periodic
boundary conditions, in  going to finite temperature, we also break 
supersymmetry. Therefore at high temperature  we expect to reduce ourselves
to a non-supersymmetric gauge theory that is presumably in the same
universality class as pure Yang-Mills theory in three dimensions and that is
confining. 

In the previous section we have seen that ${\cal{N}}=4$ super 
Yang-Mills at zero temperature is related to $AdS_5$ and it is therefore
natural to expect that ${\cal{N}}=4$ super Yang-Mills at finite temperature 
is related to the finite temperature version of $AdS_5$ discussed in 
Ref.~\cite{HAWKING}. We will see that at finite temperature we need to 
consider, at least for very large 't Hooft
coupling where supergravity is a good approximation, two classical solutions 
of the supergravity equations: the first one is $AdS_5$ that we had also at
zero temperature and that is dominating at low temperature, while the second one
is the $AdS_5$ black hole that is instead dominating at high temperature.
The high temperature case is the most interesting one because in this
case we get confinement and a mass gap.

In section~\ref{AdS} we have seen that anti De Sitter space in euclidean
uncompactified space is described by  eq.(\ref{locus2}). 
We now want to compactify the coordinate v. 
Let us restrict ourselves to $AdS_{n+1}^{+}$ where both 
$u, v >0$. The manifold in eq.(\ref{locus2})  is invariant under the action of
a group that we call Z and that acts on the coordinates as follows:
\beq
u \rightarrow \lambda^{-1} u \hspace{2cm} v \rightarrow \lambda  v
\hspace{2cm} y^{\alpha} \rightarrow y^{\alpha}
\label{zz56}
\eeq
We can construct a compactified version of $AdS_{n+1}^{+}$ by modding out the
action of the group Z. In this way one gets the manifold $X_1 \equiv 
AdS_{n+1}^{+}/ Z$. The fundamental 
domain for the action of Z on $v$ is the interval $ 1 \leq \frac{v}{b} 
\leq \lambda$. Therefore $v$ parametrizes a circle with natural coordinate
\beq
\frac{v}{b} = \lambda^{\theta/2 \pi} \hspace{2cm} 0 \leq \theta \leq 2 \pi
\label{cic56}
\eeq
The independent variables describing this compactified version of anti De 
Sitter space can be taken to be
$v$ and $\vec{y}$, while $u$ is given in terms of them through 
eq.(\ref{locus2}). The manifold $X_1$ spanned by $(v, \vec{y} )$ is 
then topologically
equivalent to $S^1 \times R^n$. The boundary of $X_1$ is instead topologically
equivalent to $S^1 \times S^{n-1}$. It is convenient to perform the change of
variables:
\beq
\frac{t}{b} = \log \frac{v}{b} - \frac{1}{2} \log \left( b^2 + r^2 \right)
\hspace{.5cm};\hspace{.5cm} r^2 \equiv \sum_{\alpha=1}^{n} y_{\alpha}^{2}
\label{newva4}
\eeq
Then the metric of $X_1$ becomes
\beq
(ds^2 )_{X_1} = dt^2\left[ 1 + \frac{r^2}{b^2}\right] + 
\frac{dr^2}{1 + \frac{r^2}{b^2}} + r^2 d \Omega_{n-1}^{2}
\label{x1}
\eeq
The previous metric is a compactified version of a solution of the eq. of 
motion given in eq.(\ref{eineq}), that has, however, also another solution. 
This is the $AdS_{n+1}$ black hole whose metric is given by:
\beq
(ds^2 )_{X_2} = dt^2 \left[ 1 + \frac{r^2}{b^2} - \frac{w_{n} M}{r^{n-2} }
\right] + 
\frac{dr^2}{1 + \frac{r^2}{b^2} - \frac{w_{n} M}{r^{n-2}} } + 
r^2 d \Omega_{n-1}^{2}
\label{x2}
\eeq
where
\beq
w_n = \frac{16 \pi G_{N}}{(n-1) \Omega_{n-1}} \hspace{2cm}
\Omega_{n-1} = Vol (S^{n-1})
\label{wn65}
\eeq 
The horizon of the black hole corresponds to the largest root of the equation:
\beq
V( r_{+} ) \equiv 1 + \frac{r^{2}_{+}}{b^2} - \frac{w_{n} M}{r^{n-2}_{+}} =0
\label{hori}
\eeq
$X_2$ is topologically equivalent to $R^2 \times S^{n-1}$ where $R^2$ 
corresponds to the variables $(r,t)$. Its boundary has the topology of
$S^1 \times S^{n-1}$. In conclusion we have two solutions of the classical
eq.(\ref{eineq}): the manifold $X_1$ corresponding to $AdS_{n+1}$ with a
compactified coordinate having the topology of $S^1 \times R^n$ and the
manifold $X_2$ corresponding to the anti De Sitter black hole having the
topology of $R^2 \times S^{n-1}$. Their boundary has in both cases the
topology of $S^1 \times S^{n-1}$ that is also the topology of Minkowski
space with compactified time and space.

We now want to show that at the horizon of the black hole there is no
singularity if $t$ is a periodic variable with period equal to:
\beq
\beta_0 \equiv \frac{1}{T} = \frac{4 \pi b^2 r_{+}}{n r_{+}^{2} + (n-2) b^2}
\label{tempe5}
\eeq
This can be easily obtained by expanding the metric around the horizon
\beq
(ds^2 )_{X_2} = V' (r_+ ) ( r - r_{+} ) dt^2 + 
\frac{dr^2}{V' (r_+ ) ( r - r_{+} )} + \dots
\label{hori876}
\eeq
where
\beq
V(r) = V' (r_+ ) ( r - r_{+} ) +  \dots \hspace{2cm} 
V' (r_+ ) = \frac{n r_{+}^{2} + (n-2) b^2}{r_{+} b^2}
\label{expame}
\eeq
By introducing the new variables:
\beq
z = \frac{2  ( r - r_{+} )^{1/2}}{ \left( V' (r_+ ) \right)^{1/2} }
\hspace{2cm} \theta = \frac{1}{2} V' (r_+ )  t
\label{var542}
\eeq
the metric in eq.(\ref{hori876}) becomes a two-dimensional flat metric in 
polar coordinates:
\beq
(ds^2 )_{X_2} = dz^2 + z^2 d \theta^2  
\label{polco}
\eeq
There is no singularity if the variable $\theta$ is periodic with period
equal to $2 \pi$. Then eq.(\ref{var542}) implies that $t$ must also be a 
periodic variable with period equal to $\beta_0$ given in eq.(\ref{tempe5}).
If we plot $\beta_0$ as a function of $r_{+}$ we see that $\beta_0$ is 
vanishing for both $r_{+} =0$ and $ \infty$ and has a maximum value equal to
$(4 \pi b)/(\sqrt{n(n-2)})$ for $r_{+} = b \sqrt{(n-2)/n}$. This means that
$\beta_0$ cannot be arbitrarily large and therefore the temperature cannot
be arbitrarily small. We will in fact see that this solution is relevant
for high temperature, while the other solution $X_1$ is relevant at low
temperature. In order to see which one of the two solutions dominates we
have to compute their classical euclidean action. In both cases it is equal to:
\beq
 I_{class} = \frac{n V_{n+1}}{8 \pi G_N b^2}
\label{claact2}
\eeq
where $V_{n+1}$ is a divergent volume. Therefore the classical action is
infinite in both cases. We compute their difference by regularizing each
of the contributions with a radius $R$ and by taking the two temperatures
connected by the condition:
\beq
\sqrt{ 1 + \frac{R^2}{b^2}} \beta_{0} (X_1 ) =
\sqrt{ 1 + \frac{R^2}{b^2} - \frac{w_n M}{b^{n-2}}} \beta_{0}  
\label{reg765}
\eeq
that relates the two periods in a coordinate invariant way. 
Therefore we have to compute:
\beq
I_2 - I_1 = \frac{n\Omega_{n-1}}{8 \pi G_N b^2}\left\{ \int_{0}^{\beta_0} dt
\int_{r_+}^{R} dr r^{n-1}  - \int_{0}^{\beta_0 (X_1)} dt
\int_{0}^{R} dr r^{n-1} \right\}
\label{diffe}
\eeq
The previous integrals can be easily computed and in the limit of the cutoff
$R \rightarrow \infty$ we get a finite result:
\beq
I_2 - I_1 \equiv \Delta I = \frac{\Omega_{n-1}}{4 G_N} 
\frac{b^2 r_{+}^{n-1} - r_{+}^{n+1}}{n r_{+}^{2} + (n-2) b^2}
\label{diffe4}
\eeq
If we interpret $\Delta I$ as the free energy in statistical mechanics 
we get that the energy
\beq
E  \equiv \frac{\partial \Delta I}{\partial \beta_0} = 
\frac{\partial \Delta I}{\partial r_{+}}\frac{\partial r_{+}}{\partial \beta_0}
= \frac{1}{w_n}\left(\frac{r_{+}^{n}}{b^2} + r_{+}^{n-2} \right) =M
\label{enem}
\eeq
is equal to the parameter $M$ that corresponds to the mass of the black hole
and that the entropy
\beq
S = \beta_0 E - \Delta I = \frac{\Omega_{n-1} r_{+}^{n-1}}{4 G_N}
\label{entro}
\eeq
is in complete agreement with the  Beckenstein-Hawking expression for the 
entropy of a black hole given by the area of the horizon in $(n-1)$
dimensions divided by $4 G_N$. In particular, if we consider the case $n=3$
and we take the anti De Sitter radius $b \rightarrow \infty$ we get the
metric of the Schwarzschild black hole given by:
\beq
(ds^2 )_S = ( 1 - \frac{r_g }{r}) dt^2 + \frac{dr^2}{(1 - \frac{r_g }{r})}+
r^2 d \Omega_{2}^{2}
\label{Schwar}
\eeq
where $r_g = w_3 M = 2 G_N M $ is the Schwarzschild radius.

For both $X_1$ and $X_2$ the radii of the boundary ($r \rightarrow \infty$)
with the topology of $S^1 \times S^{n-1}$ are given by:
\beq
\beta = \frac{r \beta_0}{ 2 \pi b} \hspace{2cm} \beta' = r \hspace{1cm}
\frac{\beta}{\beta'} = \frac{\beta_0}{2 \pi b}
\label{rad86}
\eeq
Therefore in the decompactification limit in which the topology of
$S^1 \times S^{n-1}$ becomes that of $S^1 \times R^{n-1}$, i.e. the
topology of Minkowski space with periodic euclidean time, we must take
the high temperature limit $\beta_0 \rightarrow 0$. This limit can be
obtained for both $r_{+} \rightarrow 0$ and $r_{+} \rightarrow \infty$.
We will see later on that actually the high temperature phase corresponds
to the case $r_{+} \rightarrow \infty$ because this branch is dominant with
respect to the other. In this limit corresponding also to the limit $M 
\rightarrow \infty$, as one can see from eq.(\ref{enem}), we get:
\beq
r_{+} = ( w_n M b^2 )^{1/n} \hspace{2cm} \beta_0 = \frac{4 \pi b^2}{n r_{+}}
\label{resca32}
\eeq
When $r_{+} \rightarrow \infty$ ($M \rightarrow \infty$) it is 
convenient to introduce the new variables:
\beq
r = \left( \frac{w_n M}{b^{n-2}}\right)^{1/n} \rho \hspace{2cm}
t = \left( \frac{w_n M}{b^{n-2}}\right)^{- 1/n} \tau = \frac{b}{r_{+}} \tau 
\label{res32}
\eeq
In terms of them the metric in eq.(\ref{x2}) becomes:
\beq
(ds^2 )_{X_2} = \left(\frac{\rho^2}{b^2} - \frac{b^{n-2}}{\rho^{n-2}} \right)
d \tau^2 + \frac{d \rho^2}{\frac{\rho^2}{b^2} - \frac{b^{n-2}}{\rho^{n-2}}}
+ \rho^2  \left(\frac{w_n M}{b^{n-2}}\right)^{2/n}  d \Omega_{n-1}^{2}
\label{newme42}
\eeq
Notice that, when $M \rightarrow \infty$, the radius of $S^{n-1}$ becomes
very large and the period of the variable $\tau$ becomes  equal to 
$(4 \pi b)/n$.

Both solutions $X_1$ and $X_2$ contribute to the partition function and
correlators of the gauge theory that in our case is ${\cal{N}}=4$ super
Yang-Mills at finite temperature. In general we have to sum over both of 
them:
\beq
e^{-I} \rightarrow e^{- I_1} + e^{- I_2} = e^{-I_1} \left[1 + e^{- \Delta I} 
\right] = e^{- I_2} \left[ 1 + e^{\Delta I} \right]
\label{act65}
\eeq  
From eq.(\ref{diffe4}) we see that, when $r_{+}$ is small, then $\Delta I
> 0$ and therefore $X_1$ dominates. This is the limit that
describes the low temperature phase. When instead $r_{+} \rightarrow \infty$
from the same eq. we see that $\Delta I < 0$ and therefore the solution 
$X_2$ is dominant at high temperature. One sees also that the branch at
$r_+ \rightarrow \infty$ is dominant with respect to the one $r_+ \rightarrow
0$ because at $r_+ \rightarrow 0$ $I_2 = I_1$, while at $r_+ \rightarrow 
\infty$ $I_2 < I_1$.

Following the Maldacena conjecture we expect that ${\cal{N}}=4$ super Yang-Mills
at high temperature and for large 't Hooft coupling ($\lambda \rightarrow 
\infty$)
is described by the $AdS_5$ black hole. In order to check this let us compare 
the entropy of ${\cal{N}}=4$ super Yang-Mills with that of $AdS_5$ black 
hole~\cite{KLEBA5}. The entropy of the $AdS_5$ black hole can
be obtained from eq.(\ref{entro}) for $n=4$. By rewriting it in terms of the
temperature related to $r_{+}$ through eq.(\ref{resca32}) ($\beta_0 = 1/T$) 
and introducing $V_3 = \Omega_3 b^3$ we get:
\beq
S_{BH} = \frac{\pi^2}{2}  V_3 T^3 N^2 
\label{bhent}
\eeq
where we have used eq.(\ref{rad}) and the fact that the five-dimensional Newton
constant is equal to $16 \pi G_{N}^{(5)} = (2 \pi)^7 (\alpha ')^4 
g_{s}^{2}/(b^5 \Omega_5)$ with $\Omega_5 = \pi^3$. The factor $ b^5 \Omega_5$
is the volume of $S^5$.

The entropy of ${\cal{N}}=4$ super Yang-Mills can be easily computed at weak 
coupling where it can just be obtained by counting the bosonic and fermionic 
degrees of freedom. In fact by taking into account that ${\cal{N}} =4$ super 
Yang-Mills theory has $8$ bosonic and $8$ fermionic massless degrees of 
freedom and that the entropy of each bosonic and fermionic degrees of freedom 
is  given respectively by:
\beq
S_{BOS} = \frac{2 \pi^2}{15 \cdot 3} N^2 V_3 T^3
\hspace{1cm}; \hspace{1cm}
S_{FER} = \frac{7}{8} S_{BOS} = \frac{14 \pi^2}{15 \cdot 3 \cdot 8} 
N^2 V_3 T^3
\label{entrobf}
\eeq
we get the following entropy at weak coupling
\beq
S_{YM} = \frac{2 \pi^2}{3} N^2 V_3 T^3
\label{ymentro}
\eeq 
It is equal to the entropy of the black hole in eq.(\ref{bhent}) apart from a 
numerical factor
($4/3$). The mismatch between the two results can be easily explained from
the fact that one is valid for strong coupling while the other one is valid
in perturbation theory~\cite{KLEBA6}. In general we expect the following 
behaviour of the entropy with the Yang-Mills coupling constant~\cite{KLEBA6}:
\beq
S (N g_{YM}^{2} ) = \frac{2 \pi^2}{3} N^2 V_3 T^3 f(N g_{YM}^{2})
\label{entro76}
\eeq
where $f (x)$ is a smooth function that is equal to $f(0) =1$ at weak coupling
corresponding to $x=0$ and to $f(\infty) = 3/4$ at strong coupling. The
inclusion of the first string correction~\cite{KLEBA6} gives:
\beq
f (N g^{2}_{YM} ) = \frac{3}{4} + \frac{45}{32} \zeta (3) 
( 2 g_{YM}^{2} N)^{-3/2} 
\label{strico3}
\eeq
while a recent two-loop calculation~\cite{TAY} shows that its perturbative 
expansion is:
\beq
f (N g^{2}_{YM} ) = 1 - \frac{3}{2 \pi^2}  g^{2}_{YM} N
\label{pertex2}
\eeq
These results show that the function $f$ is not a constant and are consistent
with $f$ being a monotonic function interpolating between $1$ at $N g_{YM}^{2}
=0$ and $3/4$ at $N g_{YM}^{2} = \infty$.

In the second  part of this section we consider ${\cal{N}} =4$ super Yang-Mills
at high temperature and we compute various physical quantities in this theory 
using its correspondence with IIB supergravity compactified on 
$AdS_5 \times S^5$. As we
have explained in Sect.~\ref{malda} supergravity is a good approximation to
${\cal{N}}=4$ super Yang-Mills in the strong coupling limit $N g_{YM}^{2} >>1$.
In particular in the following we will show that, by computing the Wilson 
loop and finding that it is 
proportional to the area, this theory confines. We will then look at the glue 
ball mass spectrum and
we will show that in the high temperature phase a mass gap is generated.

For computing the Wilson loop it is convenient to use the following form of 
the black hole metric:
\beq
ds^2 = (\alpha')^2 \frac{U^2}{b^2} \left[ \left( 1 - \frac{U_{T}^{4}}{U^4} 
\right) dt^2 + \sum_{i=1}^{3} (d x^i )^2 \right] + \frac{b^2 dU^2}{U^2 
\left( 1 - \frac{U_{T}^{4}}{U^4} \right)} + b^2 d\Omega_{5}^{2}
\label{metri54}
\eeq
where the variables used here are related to those used in eq.(\ref{newme42})
for $n=4$ by the equations:
\beq
\tau = \alpha' \frac{U_T}{b} t \hspace{1cm} \rho = \frac{b}{U_T} U
\label{newva2}
\eeq
In eq.(\ref{metri54}) we have also added the part of the metric 
corresponding to the sphere $S^5$.
The variable $U$ is the same as the one defined in eq.(\ref{lim}).
In these new variables the period of the periodic variable $t$ is equal to:
\beq
\beta = \frac{1}{T} = \frac{\pi b^2}{\alpha' U_T}
\label{tempe6}
\eeq
Following Refs.~\cite{MALDA2,REY} the rectangular Wilson loop in the gauge
theory can be approximated in the strong coupling limit by the value of the
minimal Nambu-Goto string action. The string has the world sheet in 
anti De Sitter space ending on the rectangular Wilson loop. The string action
in anti De Sitter space is given by:
\beq
S = \frac{1}{2 \pi \alpha'} \int d \sigma \int d \tau \sqrt{\det (G_{MN} 
\partial_{\alpha} x^M \partial_{\beta} x^N)}
\label{sact}
\eeq  
where $G_{MN}$ is the metric in eq.(\ref{metri54}). The Wilson loop is along
the variables $x_1$ and $x_2$ and we choose the static gauge, where $x_2 =
\sqrt{\alpha'} \tau$ and $x_1 = \sqrt{\alpha '} \sigma$. The finite 
temperature calculation has been carried out in Refs.~\cite{REY2,YANKI}.
For the sake of 
simplicity we consider the case in which the world sheet of the string  
depends only on the variable $U$ where in particular $U$ depends only on one 
of the world sheet variables $x_{1}$. In this case for the action
in eq.(\ref{sact}) we get the following expression:
\beq
S = \frac{X_2}{2 \pi} \int_{-R/2}^{R/2} d x \left\{ 
\frac{U^4 (\alpha')^2}{b^4}+ \left[ 1 - \frac{U_{T}^{4}}{U^4} \right]^{-1} 
\left( \frac{d U}{d x}\right)^2 \right\}^{1/2}
\label{act4}
\eeq
where $x \equiv x_1$, $X_2$ is the lenght of the "temporal" side of the 
rectangular Wilson
loop and $R$ is the distance between the test "quarks". Since the previous
Lagrangian does not explicitly depend  on $x$ the corresponding hamiltonian is
a constant of motion:
\beq
- H = \frac{U^4 (\alpha ')^2}{b^4} \left\{ \frac{U^4 (\alpha ')^2}{b^4}
+ \left[ 1 - \frac{U_{T}^{4}}{U^4} \right]^{-1} \left( \frac{d U}{d x}\right)^2
\right\}^{-1/2}  = C
\label{hami8}
\eeq
From it we get
\beq
\left(\frac{dU}{dx} \right)^2 = \frac{(\alpha')^2}{ b^4}
\left[\frac{U^4 (\alpha')^2}{C^2 b^4}-1 \right] \left[ U^4  - U_{T}^{4}
\right]
\label{dudx}
\eeq
Introducing the minimum value of $U$, that we call $U_0$, corresponding, because
of the symmetry of the problem, to $x=0$, and that is the
value for which eq.(\ref{dudx}) vanishes, we can determine the constant $C$
in terms of $U_0$:
\beq
C^2 = \frac{U_{0}^{4} (\alpha ')^2}{b^4}
\label{consta3}
\eeq
Integrating the differential equation in (\ref{dudx}) we get
\beq
\int_{0}^{X} dx = \frac{b^2}{\alpha'} \int_{U_0}^{U} \frac{dU}{\left[ 
\left(\frac{U^4}{U_{0}^{4}} -1 \right) \left( U^4 - U_{T}^{4} \right) 
\right]^{1/2}}
\label{inte43}
\eeq
Since the two test "quarks" in the 
Wilson loop are at a distance $R$ in  Minkowski space, that is
the boundary of $AdS_5$ obtained by taking the limit $U \rightarrow \infty$,
the value $X=R/2$ in the l.h.s. of eq.(\ref{inte43}) corresponds to 
$U \rightarrow
\infty$. Introducing the variable $ w = U/ U_0 $ and taking this limit in
the previous eq. we get:
\beq
\frac{R}{2} = \frac{b^2}{\alpha' U_0 }\int_{1}^{\infty} 
\frac{dw}{\sqrt{(w^4 -1)( w^4 - \left( \frac{U_T}{U_0}\right)^4 ) }}
\label{rmezzi}
\eeq
Analogously we can compute the energy corresponding to the minimal surface
that is given by:
\beq
E \equiv \frac{S}{X_2} = \frac{U_0}{\pi} \int_{1}^{\infty} dw
\frac{  w^4}{\sqrt{(w^4 -1)( w^4 - \left( 
\frac{U_T}{U_0}\right)^4 ) }}
\label{ene43}
\eeq
This quantity is divergent and can be regularized by cutting it off at a value
$U_{max}/U_0$ and subtracting to it the self-energy of the two test "quarks"  
$(U_{max} - U_T )/\pi$ corresponding to the energy of two strings stretching
up to the boundary of $AdS_5$. 
The energy is now convergent and we can take the limit $U_{max} \rightarrow
\infty$ obtaining
\beq
E = \frac{U_0}{\pi} \int_{1}^{\infty} dw \left\{ \frac{  w^4}{\sqrt{(w^4 -1)
( w^4 - \left( \frac{U_T}{U_0}\right)^4 )}  } -1 \right\} + 
\frac{U_T - U_0}{\pi}
\label{enere}
\eeq 
The previous equation can be rewritten as:
\beq
E = \frac{U_0}{\pi} \int_{1}^{infty} dw  
\left\{ \frac{\sqrt{(w^4 -1)}}{(w^4 - \left( \frac{U_T}{U_0}\right)^4 )} -1 
\right\} + 
\frac{U_T - U_0}{\pi} + \frac{U_{0}^{2} \alpha '}{2 \pi b^2} R
\label{enefi8}
\eeq
From eq.(\ref{rmezzi}) it is easy to see that $U_0 \rightarrow U_T$ when 
$R \rightarrow \infty$. In this limit all terms in eq.(\ref{enefi8}) vanish 
except 
the last one that gives a confining potential with string tension given by:
\beq
\frac{E}{R} \equiv \sigma = \frac{U_{T}^{2} \alpha '}{ 2 \pi b^2} = 
\frac{\pi b^2}{2 \alpha'} T^2
\label{strte}
\eeq  
where we have used eq.(\ref{tempe6}). Using eq.(\ref{rad2}) we get finally:
\beq
\sigma = \frac{\pi}{2} \sqrt{N g_{YM}^{2}} T^2
\label{stringte}
\eeq
In conclusion we have shown that the high temperature phase of ${\cal{N}}=4$
super Yang-Mills is a confining one with string tension given in 
eq.(\ref{stringte}).

In the following  we discuss in some detail the fact that at finite
temperature a mass gap appears. The appearence of a mass gap is usually seen 
by studying the two-point function involving for instance the composite $F^2$ 
and showing that its large distance behaviour decays exponentially with the 
distance:
\beq
< F^2 (x) F^2 (0) > \sim e^{- m |x|}
\label{massgap}
\eeq 
From this behaviour one can just read the mass $m$ of the lowest lying state 
that has the same quantum numbers as $F^2$. This provides the mass gap
of the theory. We could just compute the mass gap from the  correlator in 
eq.(\ref{massgap}), but there is a simpler way that we are going to discuss now
following Ref.~\cite{WITTEN3}.
If the Maldacena conjecture is right and therefore 
${\cal{N}}=4$ super Yang-Mills is equivalent for large values of $\lambda$ to
type IIB supergravity compactified on $AdS_5 \otimes S^5$, then the Hilbert
spaces of these two theories must be the same. 
We have seen in sect.~\ref{malda} that
the composite and gauge invariant field $F^2$ corresponds in supergravity
to the dilaton field $\phi$. In order to construct the Hilbert space in the 
supergravity approximation, that is relevant for constructing the two-point
function involving for instance two fields $F^2$ we have to consider the 
classical equation of motion of the dilaton, that is the field corresponding
to $F^2$, in the $AdS_5 \otimes S^5$ background and search for its solutions
satisfying certain boundary conditions.
Since the dilaton is massless in ten dimensions, the 
$\ell=0$ mode on $S^5$ is also massless in five dimensions. This mode does
not depend on the coordinates of $S^5$ and satisfies the classical equation:
\beq
\partial_{\mu} \left[\sqrt{g} g^{\mu \nu} \partial_{\nu} \Phi \right] =0
\label{claeq2}
\eeq 
where $g_{\mu \nu}$ is the metric of $AdS_5$ that is given in 
eq.(\ref{newme42}) for $n=4$.
We wish to look at solutions of the previous equation that are square 
integrable in the previous metric and that correspond to plane waves in
the boundary gauge theory, i.e. we take the dilaton in eq.(\ref{claeq2})
to be of the following form:
\beq
\Phi (\rho, x ) = f (\rho) \,\,e^{i k \cdot x}
\label{planewa}
\eeq
Following the procedure discussed in great detail in Ref.~\cite{JENS}
it is convenient to rescale $\tau$ by $ \tau = b^2 {\tilde{\tau}}$. 
The period of ${\tilde{\tau}}$ is equal to $\pi/b$ and 
after a rescaling of  the variables $x_i$ the  metric becomes:
\beq
\frac{ds^2}{b^2} = (\rho^2 - \frac{b^4}{\rho^2}) d{\tilde{\tau}}^2 + 
\frac{d\rho^2}{ \rho^2 - \frac{b^4}{\rho^2}} + \rho^2 \sum_{i=1}^{3} (dx^i)^2
+ d \Omega_{5}^{2}
\label{me421}
\eeq
Inserting the previous metric in the dilaton equation with a dilaton field
given in eq.(\ref{planewa}) we get:
\beq
\frac{1}{\rho} \partial_{\rho} \left[ \rho \left(\rho^4 - b^4 \right)
\partial_{\rho} f \right] = k^2 f
\label{equa32}
\eeq
Introducing the quantity $x = \rho^2$ and rescaling $x$ by $ x \rightarrow
b^2 x$ we get the following equation:
\beq
x(x^2 -1) \frac{d^2 f}{dx^2} +(3 x^2 -1 ) \frac{df}{dx} = \frac{k^2}{4 b^2} f
\label{fiequa}
\eeq
Rescaling $k \rightarrow b^2 k$ in order to have a quantity with the dimension
of a mass we get that in terms of the rescaled variable the coefficient of the
non-derivative term in eq.(\ref{fiequa}) becomes $k^2 b^2 /4$.
As a boundary condition we need to impose that the solution  be square
integrable. This means that it must satisfy the condition:
\beq
\int d \rho \sqrt{g} | f (\rho) |^2 < \infty
\label{squain}
\eeq 
Since $ \sqrt{g} d \rho = \rho^3 d \rho = \frac{1}{2} x dx$ the previous
condition implies that $f \sim x^{-a} $ with $ a>1$. Near the horizon the
metric in eq.(\ref{me421}) behaves after a rescaling of $\rho$ as:
\beq
ds^2 \sim \left( x - \frac{1}{x} \right) ( d {\tilde{\tau}} )^2 + 
\frac{dx^2}{4(x^2 -1) }
\label{rescame6}
\eeq
We introduce the variable $z$ related to $x$ through the relation:
\beq
dz  =  \frac{dx}{2\sqrt{x^2 -1}}
\label{newva31}
\eeq
that implies $ x = \cosh (2 z)$. The coordinate 
singularity appearing in the metric in eq.(\ref{rescame6}) at $x=1$ 
corresponds in the new variable to $z=0$. Then since near the horizon
\beq
x - \frac{1}{x} = \frac{\sinh^2 2z}{\cosh 2z} \sim 4 z^2
\label{coordsi}
\eeq
the metric in eq.(\ref{rescame6}), in the near horizon limit, becomes the
two-dimensional flat metric in polar coordinates 
\beq
ds^2 = dz^2 + 4 z^2 d{\tilde{\tau}}^2
\label{nearho}
\eeq
Since the function $f (\rho)$ is only a function of the radial variable $z$
and not of the angular variable ${\tilde{\tau}}$ a proper boundary condition
for $f$ is that it is smooth at the origin, i.e. $\frac{df}{dz} =0$ at $z =0$.
But, since
\beq
\frac{df}{dz} = \frac{dx}{dz} \frac{df}{dx} = 2 \sinh 2z \frac{df}{dx}
\sim 4 z \frac{df}{dx}
\label{boud78}
\eeq 
one gets that the function $f$ must be regular at the horizon $x=1$.
In conclusion we must look for square integrable solutions of the dilaton
classical equation in eq.(\ref{fiequa}) that are regular at the horizon.
We can rewrite the differential equation in (\ref{fiequa}) as follows:
\beq
y'' + \left[ \frac{1}{x} + \frac{1}{x+1} + \frac{1}{x-1} \right] y' =
\frac{p}{x(x^2 -1)} y
\label{equa52}
\eeq
where $f = y$. The eigenvalues of eq.(\ref{equa52}) has been determined 
numerically in Refs.~\cite{CSAKI,JEVICKI,ZYSKIN}. In the following we 
describe the method proposed in 
Ref.~\cite{ZYSKIN} in order to show that the spectrum of the eigenvalues of the
differential equation in (\ref{equa52}) is discrete and in particular that 
there is a mass gap. The general solution of the previous eq. can always be 
written
as a linear combination of two independent solutions that in general have
singularities at $x= 0, 1 $ and $\infty$. Therefore in general a solution 
cannot be represented as a convergent series expansion throughout the entire
physical region $1 \leq x \leq \infty$.  It is possible, however, to consider
expansions that are convergent in either of the two intervals 
$I (\infty ) \equiv \left\{ x \epsilon C | 1 < x < \infty \right\}$ and
$I ( 1 ) \equiv \left\{ x \epsilon C | 0 < x < 2 \right\}$. They overlap 
in the interval $1 < x < 2$. In the first interval $I (\infty)$ following
Ref.~\cite{ZYSKIN} the most general solution can be witten in terms of the 
two convergent expansions:
\beq
y_{1}^{(\infty)} = \frac{1}{x^2} + \sum_{n=1}^{\infty} a_{n}^{(\infty)} 
x^{-2 -n}
\label{y1infty}
\eeq
and
\beq
y_{2}^{(\infty)} = \frac{p^2}{2} \log (x) y_{1}^{(\infty)}  + 
\sum_{n=1}^{\infty} b_{n}^{(\infty)} x^{ -n}
\label{y2infty}
\eeq 
while in the interval $I(1)$ can be written in terms of the following two
other convergent series:
\beq
y_{1}^{(1)} =  1  + \sum_{n=1}^{\infty} a_{n}^{(1)} (x-1)^{n}
\label{y11}
\eeq
and
\beq
y_{2}^{(1)} =  \log (x-1) y_{1}^{(1)}  + \sum_{n=1}^{\infty} b_{n}^{(1)} 
(x-1)^{n}
\label{y21}
\eeq
The expansion coefficients can be determined for any value of $p$ by recursion
from the differential eq.(\ref{equa52}). In general, however, the previous
solutions or any combination of them will not satisfy both boundary conditions
that we have discussed above. For certain values of $p$ it turns out that there
exists a solution that is simultaneously proportional to $ y_{1}^{(\infty)}$
and to $ y_{1}^{(1)}$. The condition for this to happen is that their Wronskian
vanishes:
\beq
W(p,x) \equiv \left( \begin{array}{cc}   
             y_{1}^{(1)}   &    y_{1}^{(\infty)} \\
            \frac{d y_{1}^{(1)}}{dx}   &    \frac{ d y_{1}^{(\infty)}}{dx}
\end{array} \right) =0
\label{wronski}
\eeq
For $1< x < 2$ both series are convergent and the Wronskian can be computed
and can be seen to have the following form:
\beq
W(p,x) = \frac{r(p)}{x (x-1)(x+1)} 
\label{wro98}
\eeq
The function $r(p)$ can also be computed to any desired accuracy.
The spectrum of $p$ is determined by the zeroes of $r(p)$. It can be seen
that there is no positive or zero eigenvalue of the differential equation
in eq.(\ref{equa52}). Therefore a mass gap is generated together with a discrete
spectrum. In particular the eigenvalue spectrum can be approximately computed
using the WKB approximation and one gets~\cite{CSAKI,MINAHAN}:
\beq
M^2 = - k^2  = 8 \pi \left[\Gamma (3/4) \right]^4 T^2  n (n+1)  
\label{wkb}
\eeq
where $n$ is an arbitrary positive integer.

In this section we have studied the behaviour of ${\cal{N}}=4$ super Yang-Mills
at high temperature. Remembering that finite temperature means that one 
direction (the euclidean time one) is compactified along a circle of radius
equal to $1/(2 \pi T)$, then the radius becomes very small at high temperature.
This means that at high temperature the theory becomes effectively 
three-dimensional and therefore studying the original four-dimensional theory
at high temperature corresponds essentially to study a three-dimensional theory
in which supersymmetry is broken and in which we expect that both
the fermions and scalars get a mass of the order of the temperature. Because 
of this it is then natural to think that this theory reduces to a theory 
of pure Yang-Mills in three dimensions. The relation between the four and the
three-dimensional coupling constants can be found by expanding the D $3$-brane
effective Born-Infeld action and keeping only the kinetic term for the gauge
field as we have done in eq.(\ref{biexp}). In this way for $p=3$ one gets 
the Yang-Mills action in four dimensions with $g^{2}_{YM4} = 4 \pi g_s$. Then
remembering that the time direction is compactified with radius equal to
$1/(2 \pi T)$ we get the Yang-Mills action in three dimensions with coupling
constant equal to:
\beq
g_{YM3}^{2} = g_{YM4}^{2} T 
\label{YM3}
\eeq   
We have seen that the use of the supergravity approximation is allowed 
only in the strong coupling limit  where $N g^{2}_{YM4} \equiv \lambda >> 1$,
while the three-dimensional Yang-Mills scale $N g_{YM3}^{2}$ is obtained in the
limit in which $T \rightarrow \infty$ and  $N g^{2}_{YM4} \rightarrow 0$.
In order to study this limit we have to go away from the supergravity
approximation and take into account the tree diagrams of string theory. But 
this is unfortunately beyond our reach at present.

\sect{$D=4$ Yang-Mills from the M-theory $5$-brane}
\label{4dimym}

In the previous section we have seen how, starting from the ten-dimensional
non extremal D $3$-brane, one can describe strongly coupled ${\cal{N}} =4$
super Yang-Mills 
at high temperature or alternatively a theory that is presumably in the same
universality class of three-dimensional Yang-Mills theory. 
In this section we discuss a suggestion, made by Witten~\cite{WITTEN3}, on how
to extend the previous procedure from three to four-dimensional Yang-Mills 
theory. The starting point in this case is not the D $3$-brane of the 
ten-dimensional type IIB string theory, but the $5$-brane of the 
eleven-dimensional M-theory. This solution has in the near horizon limit a 
metric corresponding to the manifold $AdS_7 \otimes S^4$ with the two radii 
given by:
\beq
R_{AdS_7} \equiv b = 2 \ell_{p} ( \pi N )^{1/3} \hspace{2cm} 
R_{S^4} \equiv L =  \frac{b}{2} =  \ell_{p} ( \pi N )^{1/3} 
\label{rad74}
\eeq
where the $11$-dimensional Planck lenght $\ell_{p}$ is related to the
$11$-dimensional gravitational constant by  $ 2 \kappa_{11}^{2}
= (2 \pi)^8 \ell_{p}^{9}$. Following the notation of Ref.~\cite{HASHIOZ} the
$11$-dimensional metric of the non-extremal $5$-brane in the near horizon
limit is given by:
\beq
(ds_{11})^2 = \frac{4 L^2}{y^2} 
\frac{dy^2}{1 - \left(\frac{y_{0}}{y}\right)^{6}} + L^2 d \Omega_{4}^{2} +
\frac{y^2}{L^2} \left[ \left(1 - \frac{y_{0}^{6}}{y^6} \right) dt^2 +
\sum_{i=1}^{5} dx_{i}^{2}  \right]
\label{11metri}
\eeq
The previous variables $y$ and $t$ are related to the variables $\rho$
and $\tau$ used in eq.(\ref{newme42}) by the relations:
\beq
y = \frac{y_0}{b} \rho \hspace{1cm} t = \frac{L}{y_0} \tau
\label{chava6}
\eeq
while the parameters $y_0$ and $L$ are given in terms of the temperature and 
of the $11$-dimensional gravitational constant by:
\beq
y_0 = \frac{4 \pi}{3} T L^2 \hspace{2cm} L^9 = 
\frac{\kappa_{11}^{2} N^3}{2^7 \pi^5}
\label{tempe}
\eeq
If we compactify one the $11$ dimensions belonging to the world volume of the
$M5$-brane we obtain a $10$-dimensional theory
in which the original M $5$-brane becomes the D $4$-brane of type IIA string 
theory. The dilaton and the $10$-dimensional metric of the D $4$-brane can be 
obtained using the formula:
\beq
(ds_{11})^2 = e^{4 \phi/3} \left( d x^5 + A_{\mu} dx^{\mu} \right)^2 + 
e^{-2 \phi /3} (ds_{10})^2
\label{1110}
\eeq 
where what is usually called the $11$th dimension is here the $5$th direction.

Comparing eq.(\ref{1110}) with the $11$-th dimensional metric in 
eq.(\ref{11metri}) we get that the dilaton is given by:
\beq
e^{\phi} = g_s \left(\frac{y}{L} \right)^{3/2}
\label{dila48}
\eeq
and the ten-dimensional metric by:
\[
(ds_{10})^2 = \frac{y}{L} (ds_{11})^2 g_{s}^{2/3} =
\]
\beq
=\frac{4 L}{y} 
\frac{dy^2}{1 - \left(\frac{y_{0}}{y}\right)^{6}} + L\, y \,d \Omega_{4}^{2} +
\frac{y^3}{L^3} \left[ \left(1 - \frac{y_{0}^{6}}{y^6} \right) dt^2 +
\sum_{i=1}^{4} dx_{i}^{2}  \right]
\label{10metri}
\eeq
Notice that the dependence on the string coupling constant in the second term
of the previous equation disappears in the third term because we use variables
that are $10$-dimensional. Both $L$ and $y_0$ in eq.(\ref{10metri}) are now
expressed in $10$-dimensional units. Remembering that a lenght 
$L^{(10)}$ in $10$-dimensional units is related to a lenght $L^{(11)} $ in 
$11$-dimensional ones  through the formula :  $L^{(10)} =  g_{s}^{1/3} 
L^{(11)}$ we get that the quantity $L$  in eq.(\ref{10metri}) is 
given by:
\beq
L = \sqrt{\alpha'} ( \pi g_s N )^{1/3}
\label{newL}
\eeq
where $\sqrt{\alpha '} \equiv \ell_{p}^{(11)}$ is equal to the $11$-dimensional
Planck constant in $11$-dimensional units.

A system of $N$  D $p$-branes of type IIA theory is described at low energy by 
the non-abelian version of the Born-Infeld action given by:
\[
S_{BI} =  \tau_{p}^{(0)} \,\, STr \int d^{p+1} \xi e^{- \phi} 
\sqrt{ \det \left( 
G_{\alpha \beta} + B_{\alpha \beta} + 2 \pi \alpha ' F_{\alpha \beta} \right)}
+
\]
\beq
+ \frac{1}{\sqrt{2} \kappa_{10}^{(0)}} \int_{p+1} \sum \mu_{p-2n} A^{(p+1 -2n)} 
STr e^{F/(2 \pi)}
\label{boinfe}
\eeq
It contains external NS-NS and R-R fields that are normalized in such a way
that the Lagrangian of the bulk theory is given by:
\beq
S_{bulk} = \frac{1}{2 (\kappa_{10}^{(0)})^{2}} \int d^{10} x \sqrt{-G} \left\{
e^{- 2 \phi} \left[ R + 4 G^{\mu \nu} \partial_{\mu} \phi 
\partial_{\nu} \phi - \frac{1}{12} (H_{3})^2 \right] - \frac{1}{2 (p+2)!}
H_{p+2}^2 \right\} 
\label{bulkac}
\eeq
where
\beq
\tau_{p} \equiv \frac{\tau_{p}^{(0)}}{g_s} =
\frac{(2 \pi \sqrt{\alpha'})^{1-p}}{2 \pi \alpha' g_s}~~~~~~~;~~~~~~~
\mu_p = \sqrt{2 \pi} (2 \pi \sqrt{\alpha'})^{3-p}
\hspace{1cm} 
\label{consta}
\eeq
Remember also that $2 \kappa_{10}^{2} \equiv 2 (\kappa_{10}^{(0)})^2 g_{s}^{2}=
 (2 \pi)^7 (\alpha ')^4 g_{s}^{2}$.
We keep in the Born-Infeld action only the gauge field and the zero component 
of the $1$-form potential. Then using the
formulas given in eq.(\ref{consta}) and compactifying the time (temperature) 
direction we can rewrite eq.(\ref{boinfe}) for the case of a D 
$4$-brane as follows:
\beq
S_{BI} = \frac{\tau_{4}}{T } STr \int d^4 x 
\sqrt{\det\left( G_{\alpha \beta} + 2 \pi \alpha ' F_{\alpha \beta}  \right)} + 
 \frac{1}{2}
( 2 \pi \alpha ' )^2 \tau_{4}^{(0)} \int A^{(1)} STr \left( F_{2}^{2} \right)
\label{sborin}
\eeq
Using the relation:
\beq
\int A^{(1)} STr \left( F_{2}^{2} \right) = \frac{1}{4 T} \int d^4 x 
\epsilon^{\alpha \beta \gamma \delta \eta} A_{\alpha} Tr \left(F_{\beta \gamma}
F_{\delta \eta} \right) 
\label{equa48}
\eeq
keeping only the time component of the R-R field $A_{\alpha}$ and expanding the
Born-Infeld action restricting ourselves only to the quadratic term in F we get:
\beq
S_{BI} = \frac{\tau_{4}}{T} (2 \pi \alpha')^2  \int d^4 x 
\left\{   \frac{1}{4} Tr (F^2 )  + \frac{1}{8} g_s A_0 Tr 
\left( \epsilon^{\mu \nu \rho \sigma} F_{\mu \nu } F_{\rho \sigma} \right)
\right\}
\label{sbin}
\eeq
Using the equation:
\beq
\frac{ \tau_{4} ( 2 \pi \alpha ')^2}{T } = \frac{1}{(2 \pi)^2 
T \sqrt{\alpha'} g_s}
\label{eqsra}
\eeq
and introducing a parameter $\lambda$ through the equation:
\beq
2 \pi \sqrt{\alpha'} g_s = \frac{\lambda}{TN}
\label{parala}
\eeq
we can rewrite eq.(\ref{sbin}) as follows:
\beq
S = \frac{N}{2 \pi \lambda} \int d^4 x \left[ 
\frac{1}{4} Tr (F^2 ) + \frac{1}{8} g_s A_0 Tr 
\left(\epsilon^{\mu \nu \rho \sigma}
F_{\mu \nu} F_{\rho \sigma} \right) \right]
\label{acti876}
\eeq
From eq.(\ref{acti876}) we can read the value of the Yang-Mills coupling
constant and connect it with the radius $R_{5}^{(10)} $ of the 
$5$th direction in $10$-dimensional units. We get:
\beq
\frac{N g_{YM}^{2}}{2 \pi} = \lambda  = 2 \pi R_1 TN \hspace{1cm} 
R_{5}^{(10)} \equiv R_1 =  \sqrt{\alpha'} g_s 
\hspace{1cm} \frac{1}{T} \equiv 2 \pi R_2 
\label{frac453}
\eeq
or in other words:
\beq
g_{YM}^{2} = (2 \pi R_1) (2 \pi T) = \frac{2 \pi R_1}{R_2} 
\label{ymcoupli8}
\eeq
To summarize we started with the M-theory $5$-brane having a
six-dimensional world volume. We have then compactified two of the six
directions on two radii. The first one, that we called $R_1$, corresponds in
going from the $11$-dimensional M-theory to the $10$-dimensional type IIA
string theory and because of this compactification the M-theory $5$-brane 
becomes the 
$10$-dimensional  D $4$-brane of type IIA theory. The second radius $R_2$ that
from the point of view of the $5$-dimensional theory, corresponding to the world
volume of the D $4$-brane, is related to the temperature  which is also
the temperature of the non-extremal black hole solution, has been 
introduced in
order to reduce ourselves to a $4$-dimensional theory gauge theory that we are
interested to study. In particular in the temperature direction we are free to
choose antiperiodic 
boundary conditions for the fermions of the theory and correspondently they 
will have masses equal to $(2n+1)/R_2$ ($n$ is an integer) that become very
big in the high temperature limit ($R_2 \rightarrow 0$). Also the scalars of
the theory will get a mass of the order $g^{2}_{YM} /R_2$ and they will also 
become
very massive in the high temperature limit. Therefore in this limit we will be
left with only the gauge field and it looks plausible that such a theory
is in the same universality class as pure Yang-Mills theory. 

In the second part of this section we will show how to compute various 
observables
as for instance the Wilson loop~\cite{YANKI} and the topological 
susceptibility~\cite{HASHIOZ}  
in the previously defined four-dimensional theory. Let us start
computing the topological susceptibility.
In eq.(\ref{acti876}) we see that the Yang-Mills topological charge density
is coupled to the time component $g_s A_0 \equiv h $ of the R-R field $A_{\mu}$:
\beq
h \rightarrow {\tilde{O}}_4 = \frac{N}{2 \pi \lambda} \cdot \frac{1}{8}
\epsilon^{\mu \nu \rho \sigma} Tr \left( F_{\mu \nu} F_{\rho \sigma} \right)
\label{tildeo4}
\eeq
where $\mu, \nu, \rho $ and $\sigma$ are all four-dimensional indices.
 Therefore
in order to compute correlators involving the topological charge density as
for instance the topological susceptibility, that is related to the two-point 
function
with two operators ${\tilde{O}}_4$, we must look at the classical eq. of motion
for $A_0$ that follows from the type IIA supergravity Lagrangian. The relevant
term of the type IIA supergravity is the kinetic term for $A_{\mu}$, namely
\beq
\frac{1}{2 (\kappa_{10}^{(0)})^{2}} \frac{1}{4} \int d^{10} x 
\sqrt{g} F_{\mu \nu}
F^{\mu \nu} = \frac{1}{2 (\kappa_{10}^{(0)} )^{2}} \int d^{10} x \sqrt{g} \frac{1}{2}
g^{\mu \nu} g^{00} \partial_{\mu} A_{0} \partial_{\nu} A_{0}
\label{ff4}
\eeq 
The function $h \equiv g_s A_{0}$ is determined by solving the classical eq. 
that follows from the previous action:
\beq
\partial_{\mu} \left[ \sqrt{g} g^{00} g^{\mu \nu} \partial_{\nu} h \right] =0
\label{equa31}
\eeq
We can assume that $h$ is only a function of $y$. In fact, since we will see
that the contribution to the various correlators comes from a boundary term, 
the dependence on the other variables will be irrelevant because they will
never contribute to a total divergence. 
Remembering that the background metric following from eq.(\ref{10metri}) is
given by:
\beq
g^{00} = \left(\frac{L}{y}\right)^3 
\frac{1}{1 - \left( \frac{y}{y_0} \right)^6} \hspace{1cm}
g^{yy} = \frac{y}{4L} \left[ 1 - \left( \frac{y}{y_0} \right)^6  \right]
\hspace{1cm} \sqrt{g} = 2 \left(\frac{y}{L} \right)^9
\label{me62}
\eeq
we get the following eq. of motion:
\beq
\partial_{y} \left[ \left( \frac{y}{L} \right)^7 \partial_{y} h  \right] =0
\label{eq31}
\eeq
Integrating it with the two  boundary conditions:
\beq
\lim_{y \rightarrow \infty} h(y) = h^{\infty} \hspace{2cm} h(y_0 ) =0
\label{boundco}
\eeq
we obtain the following solution~\cite{WITTEN8,HASHIOZ}:
\beq
h (y) = h^{\infty} \left[ 1 - \left(\frac{y_0}{y} \right)^6 \right]
\label{sol96}
\eeq
When we insert it in the action in eq.(\ref{ff4}) we are left with
only a surface term that cannot be neglected and is equal to:
\[
S_{class.}= \frac{1}{2 \kappa_{10}^{2}} \int d^{10} x \partial_{y}
\left[\frac{\sqrt{g}}{2} g^{yy} g^{00} h \partial_{y} h \right]=
\]
\beq
=\frac{1}{2 \kappa_{10}^{2}}\int d \Omega_4 L^4 \int d^4x \int dt \left( 
\frac{\sqrt{g}}{2} g^{00} g^{yy} \right) h \partial_{y} h |_{\infty}
\label{cla6}
\eeq
Using eqs.(\ref{me62}) and (\ref{sol96}) we get:
\beq
h \partial_{y} h |_{\infty} = 6 ( h^{\infty})^2 \frac{y_{0}^{6}}{y^7}
\hspace{2cm} \frac{1}{2} \sqrt{g} g^{yy} g^{00}|_{\infty} = \frac{1}{4} 
\left(\frac{y}{L} \right)^7
\label{rel94}
\eeq
Inserting them in eq.(\ref{cla6}) we see that the dependence on $y$ cancels
out and we get
\beq
S_{class.} = \frac{1}{2 \kappa_{10}^{2} } ( h^{\infty} )^2 
\frac{3 y_{0}^{6}}{2 L^7} \int d \Omega_4 L^4 \int d^4x \int dt 
\label{clsa9}
\eeq 
Using the  following expressions
\beq
\int d\Omega_4 = \frac{8 \pi^2}{3} \hspace{2cm} \int d^4 x \equiv V_4 
\hspace{1cm} \int dt = \frac{1}{T}
\label{rex}
\eeq
eq.(\ref{clsa9}) becomes
\beq
S_{class.} = V_4 \frac{2 \pi^2 y_{0}^{6}}{\kappa_{10}^{2} L^3 T} ( h^{\infty})^2
\label{clra}
\eeq
The correlator involving two operators ${\tilde{O}}_4$ is obtained 
by differentiating twice $e^{- S_{class.}}$ with
respect to $h^{\infty}$ and eliminating the volume factor. We get:
\beq
\int d^4 x < {\tilde{O}}_{4} (x)  {\tilde{O}}_{4} (0)> = 
\frac{4 \pi^2 y_{0}^{6}}{\kappa_{10}^{2} L^3 T} = 
\frac{N^2 T^4 \pi^3 2^7 \lambda}{3^6}
\label{top}
\eeq
where we have used eqs.(\ref{tempe}) and the relation $\kappa_{10}^{2} =
\kappa_{11}^{2} NT/\lambda $ that follows from eq.(\ref{parala}). Finally the 
topological susceptibility is given by~\cite{HASHIOZ}:
\beq
\chi_{t} \equiv  \int d^4 x \left( \frac{\lambda}{2 \pi N} \right)^2 \frac{1}{4}
< {\tilde{O}}_{4} (x)  {\tilde{O}}_{4} (0)> = \frac{8 \lambda^3 T^4 \pi}{3^6}
\label{topo1}
\eeq
We can use the previous results to determine the behaviour of the vacuum energy
of a gauge theory in terms of the vacuum $\theta$ parameter. In particular
in the supergravity approximation one can show~\cite{WITTEN8} that the vacuum 
energy behaves precisely as given in eq.(\ref{ethet26}) according to the 
large $N$ considerations discussed in sect.~\ref{u1pro}. We have seen above
that the Born-Infeld action gives a term of the following form 
(see eq.(\ref{sbin})):
\beq
\int_V A \wedge Tr \left( F \wedge F \right)
\label{sbin65}
\eeq
where $V = S^1 \times R^4$. The previous equation implies that we can 
introduce a $\theta$ parameter in the four-dimensional gauge theory by
requiring that the integral of the abelian vector field $A$ along the
compactified direction be nonzero and equal to
\beq
\int_{S^1} A = \theta + 2 \pi k
\label{theta4}
\eeq
where $A$ is the value of the abelian vector field in Minkowski space 
corresponding to the limit $y \rightarrow \infty$ and we have taken care of
the fact that $\theta$ is an angular variable by extracting from it the factor
$2 \pi k$. The vacuum energy of the four-dimensional gauge theory can then be
computed by proceeding exactly as in the calculation of the topological 
susceptibility and obtaining:
\beq
E (\theta) = \frac{\chi_t}{2} Min_{k} ( \theta + 2 \pi k)^2
\label{vacene4}
\eeq
in perfect agreement with eq.(\ref{ethet3}) obtained in the framework of the
large $N$ expansion after having used eq.(\ref{topo}).  

We now turn our attention to the Wilson loop, we 
show that it is proportional to the area and from it we extract the string 
tension. 
The calculation is very similar to the one we have 
done in sect.~\ref{fintemp} for ${\cal{N}}=4$ super Yang-Mills at finite
temperature. 
One starts with the string action  in eq.(\ref{sact}) in the metric 
given in eq.(\ref{10metri}). Choosing the static gauge where
$x_1 \equiv x = \sqrt{\alpha'} \sigma$, $x_2 = \sqrt{\alpha '} \tau$,
assuming that only the anti de Sitter variable $y$ is a function of $x$
and remembering that (see eq.(\ref{10metri}) ):
\beq
G_{xx} = G_{x_2 x_2} = \left(\frac{y}{L} \right)^3 \hspace{2cm}
G_{yy} =  \frac{4L}{y}\left[ 1 - \left(\frac{y_0}{y} \right)^6 \right]^{-1}
\label{me43}
\eeq
we get the following expression for the string action:
\beq
S = \frac{X_2}{2 \pi \alpha '} \int_{-R/2}^{R/2} d x \left\{
\left(\frac{U}{L^2}\right)^3 + \frac{1}{L^2} \left[ 1 - \left( \frac{y_0^{2}}{U}
\right)^3 \right]^{-1} \left( \frac{d U}{d x}\right)^2 \right\}^{1/2}
\label{act5}
\eeq
where we have defined $y \equiv U^2$. Also in this case the hamiltonian is
a constant of motion implying that:
\beq
- H = \left(\frac{U}{L^2}\right)^3 \left\{ \left( \frac{U}{L^2} \right)^3
+  \frac{1}{L^2} \left[ 1 - \left( \frac{y_{0}^2}{U} \right)^3 \right]^{-1} 
\left( \frac{d U}{d x}\right)^2 \right\}^{-1/2}  = C
\label{hami9}
\eeq
From it we get
\beq
\left(\frac{dU}{dx} \right)^2 = L^2 \left[ \left(\frac{U}{L^2} \right)^3 - 
\left( \frac{y_{0}^2}{U} \right)^3 \right] 
\left[\left(\frac{U}{U_0} \right)^3 -1 \right]
\label{dudx2}
\eeq
where $U_0$ is the minimum value of $U$ corresponding to $C^2 = (U_0 )^3 / L^6$.
Since for symmetry reason $U= U_0$ corresponds to $x=0$ by integrating the
previous equation we get:
\beq
x = \frac{U_0}{L} \left( \frac{L^2}{U_0}\right)^{3/2} 
\int_{1}^{U/U_0} \frac{dw}{\left[ 
\left(w^3 -1 \right) \left( w^3 - \left(\frac{y_{0}^{2}}{U_0} \right)^3 \right) 
\right]^{1/2}}
\label{inte46}
\eeq
where we have introduced  the variable $ w = U/ U_0 $. If we go to the boundary,
corresponding to sending  $U \rightarrow \infty$ and correspondently 
$x \rightarrow R/2$ we get:
\beq
\frac{R}{2} = \frac{U_0}{L} \left( \frac{L^2}{U_0}\right)^{3/2} 
\int_{1}^{\infty} \frac{dw}{\left[ 
\left(w^3 -1 \right) \left( w^3 - \left(\frac{y_{0}^{2}}{U_0} \right)^3 \right) 
\right]^{1/2}}
\label{rmezzi2}
\eeq
Analogously we can compute the energy corresponding to the minimal surface
that is given by:
\beq
E \equiv \frac{S}{X_2} = \frac{U_0}{\pi\alpha ' L} \int_{1}^{\infty} dw
\frac{ w^3}{\sqrt{(w^3 -1)( w^4 - \left( 
\frac{y_{0}^{2}}{U_0}\right)^3 ) }}
\label{ene44}
\eeq
Also in this case the energy is divergent. It can be regularized by cutting 
off the integral at $U_{max}/U_0$ and subtracting
the quantity $(U_{max} - U_T )/(\pi \alpha' L)$ where $U_T \equiv y_{0}^{2}$. 
With the previous subtraction the integral in eq.(\ref{ene44}) becomes 
convergent and one can integrate up to infinity getting:
\beq
E = \frac{U_0}{\pi \alpha' L} \int_{1}^{\infty} dw \left\{ 
\frac{  w^3}{\sqrt{(w^3 -1) ( w^3 - \left( \frac{y_{0}^{2}}{U_0}\right)^3 )}
} -1 \right\} + \frac{U_T - U_0}{\alpha' \pi L}
\label{enere1}
\eeq 
In conclusion we arrive at the two equations:
\beq
R = \frac{2L^2}{U_{0}^{1/2}}\int_{1}^{\infty} \frac{dw}{\sqrt{(w^3 -1) ( w^3 -
\left(\frac{y_{0}^{2}}{U_0} \right)^3)}}
\label{erre}
\eeq
and
\beq
E = \frac{U_0}{\pi \alpha ' L } \int_{1}^{\infty} dw \left\{ 
\frac{w^3}{\sqrt{(w^3 -1 ) (w^3 - \left(\frac{y_{0}^{2}}{U_0} \right)^3 }}
-1 \right\} + \frac{U_T - U_0}{\pi \alpha ' L}
\label{eee}
\eeq
From eq.(\ref{erre}) we see that when $R \rightarrow \infty$ then $U_0 
\rightarrow U_T$. On the other hand eq.(\ref{eee}) can be rewritten in the 
form:
\beq
E = \frac{U_0}{\pi \alpha ' L } \int_{1}^{\infty} dw  \left\{ 
\left[ \frac{  w^3 -1 }{w^3-  \left( \frac{y_{0}^{2}}{U_0}\right)^3} 
\right]^{1/2}  -1 \right\} + 
\frac{U_T - U_0}{\pi \alpha' L} + \frac{U_{0}^{3/2}}{2 \pi \alpha' L^3} R
\label{enefi9}
\eeq
Then for large $R$ we get a confining potential with string tension given by:
\beq
\frac{E}{R} \equiv \sigma = \frac{ (U_T )^{3/2}}{ 2 \pi \alpha ' L^3} 
\label{strte3}
\eeq  
Using the first eq. in (\ref{tempe}), eq.(\ref{newL}), the first eq. in 
(\ref{frac453}) and eq.(\ref{parala}) we arrive at:
\beq
\sigma = \frac{8 \pi}{27} ( g_{YM}^{2} N ) T^2
\label{sig6}
\eeq
In this section we have described Witten's proposal for studying Yang-Mills
theory starting from the M-theory $5$-brane. In particular we have computed 
several observables in the strong coupling limit of the gauge theory where
the supergravity approximation can be applied. In order to understand 
large $N$ gauge theories one would like, however, to continue the previous
results from strong  to weak coupling and to show that there is no
other singularity except the one obtained when $N g^{2}_{YM} \rightarrow 0$, 
where we expect to recover the asymptotic freedom behaviour of gauge theories
for the various observables.
This is, however, at the moment a difficult problem to solve and some
new idea seems to be needed.
\vskip0.5cm
\noindent
{\large \bf Acknowledgements}
\vskip 0.5cm
\noindent
I thank M. Frau, T. Harmark, A. Lerda, A. Liccardo, R. Marotta, N. Obers, J.L.
Petersen, R. Russo and R. Szabo for many useful discussions about the subject
of these lectures.

\end{document}